\documentclass{imsart}

\RequirePackage[OT1]{fontenc}
\RequirePackage{amsthm,amsmath}
\RequirePackage[numbers]{natbib}
\RequirePackage[colorlinks,citecolor=blue,urlcolor=blue]{hyperref}
\usepackage{amssymb}
\usepackage{graphicx}
\usepackage{bbm}
\usepackage{indentfirst}
\usepackage{tikz}
\usetikzlibrary{positioning,decorations.pathreplacing,shapes}
\usepackage[english]{babel}
\usepackage{microtype}
\usepackage{amsmath}

\startlocaldefs
\numberwithin{equation}{section}
\theoremstyle{plain}

\endlocaldefs

\begin{document}

\begin{frontmatter}
\title{A Review of Dynamic Network Models with Latent Variables \thanksref{t5}}
\runtitle{Dynamic Network Models with Latent Variables}
\begin{aug}
	\author{\fnms{Bomin} \snm{Kim}\thanksref{t1,t4}},
	\author{\fnms{Kevin} \snm{Lee} \thanksref{t2,t4}},
	\author{\fnms{Lingzhou} \snm{Xue}\thanksref{t1}},
	\and
	\author{\fnms{Xiaoyue} \snm{Niu}\thanksref{t1,t3}}
	\address[t1] {Department of Statistics, The Pennsylvania State University}
	\address[t2] {Department of Statistics,	Western Michigan University}

\thankstext{t5}{This research is partially supported by the National Science Foundation grants DMS 1505256, CISE 1320219, and the National Institute of Health grant R01 AI36664-01}	
\thankstext{t4}{These authors contributed equally to this work.}
\thankstext{t3}{Corresponding author: xiaoyue@psu.edu}

\runauthor{B. Kim et al.}


\end{aug}

\begin{abstract}
We present a selective review of statistical modeling of dynamic networks. We focus on models with latent variables, specifically, the latent space models and the latent class models (or stochastic blockmodels), which investigate both the observed features and the unobserved structure of networks. We begin with an overview of the static models, and then we introduce the dynamic extensions. For each dynamic model, we also discuss its applications that have been studied in the literature, with the data source listed in Appendix. Based on the review, we summarize a list of open problems and challenges in dynamic network modeling with latent variables.
\end{abstract}

\tableofcontents
\end{frontmatter}

\section{Introduction}

As statistical modeling of network data has been posited as a major topic of interest in diverse areas of study, there have been an increasing number of books \cite{kolaczyk2009statistical,newman2011structure} and survey papers on random graphs and network models \cite{desmarais2017statistical,goldenberg2010survey,snijders2011statistical,vivar2012models,ward2011network,matias2014modeling}. Those existing surveys provide a comprehensive overview of the historical development of statistical network modeling, including the summary of network models that are not latent variable models---e.g., Exponential random graph models (ERGMs), the quadratic assignment procedure (QAP), stochastic actor oriented models (SAOMs)---as well as the latent variable models---e.g., statist latent space models (LSM) and the stochastic blockmodels (SBM). The key idea behind introducing latent variables into network analysis is to capture various forms of dependence between edges and get conditional independence in the error terms, which is one of the most challenging parts of network modeling. Considering the difference between latent variable models and the rest in the network literature, we aim to provide in-depth information on the unobserved or unmeasured structure of networks by presenting a selective review on dynamic network models with latent variables. This area has undergone significant developments in recent years, with emphasis on two classes of models, the latent space models (LSM) and the stochastic blockmodels (SBM). \\

The latent space models (LSM) \cite{hoff2002latent} assume that nodes are positioned in an $R$-dimensional latent space, and they tend to create edges to others that are closer in their latent positions. Due to this simple geometry-based assumption, the latent space models have an advantage in providing the useful visualization and interpretation of network or relational data, and thus have been widely used in numerous fields of study \cite{chang2009relational,fletcher2011social,henry2016modeling,hoff2004modeling,krafft2012topic,sweet2011modeling,sweet2013hierarchical,ward2007disputes}. \\

The stochastic blockmodels (SBM) assume that the nodes of the network are partitioned into several unobserved (latent) classes (or blocks). The framework is first introduced by Holland et al. \cite{holland-etal-1983} which focus on the case of \emph{a} \emph{priori} specified blocks, where the membership of nodes are known or assumed, and the goal is to estimate a matrix of edge probabilities. A statistical approach to \emph{a} \emph{posteriori} block modeling for networks is introduced by \cite{snijders-1997} and \cite{nowicki-2001}, where the objective is to estimate the matrix of edge probabilities and the memberships simultaneously. Since then, the communities found in the stochastic blockmodels have been interpreted meaningfully in many research fields. For example, in citation and collaboration networks, such communities could be interpreted as scientific disciplines \cite{ji-jin-2014,newman-2004}, while the communities in food web networks could be interpreted as ecological subsystems \cite{girvan-newman-2002}.\\

As dynamic network analysis---the study of networks that evolve over time--- has become an emergent scientific field in the last decade, there has been a growing number of dynamic network models that incorporate latent space model or stochastic blockmodel framework. Since they are relatively new, to the best of our knowledge, there has not been any attempt to describe the progress in dynamic latent variable models. In this work, therefore, we outline some of the prominent dynamic latent space models and summarize their interconnections, and also describe the recent approaches in dynamic stochastic blockmodels.\\

We organize the review in the following way. For each class of models (LSM and SBM), we first introduce a family of static models as the background information, then describe the dynamic models motivated by the static ones, and discuss the applications to real-world data. In Section \ref{sec:discussion}, we present two diagrams summarizing the relationships of all the models and mention some open problems and remaining challenges in dynamic network models with latent variables. We also provide a list of available data sources that have been used in past literature in the Appendix.

\section{Notations}
Throughout the paper, we consider the set of nodes $\mathcal{N}$ labeled as $1, \ldots, N$, and assume the edges represented by an $N \times N$ adjacency matrix ${Y} = (y_{ij})_{(i,j) \in \mathcal{N}}$, where $y_{ij}$ being the edge value between $i$ and $j$ which could be binary, discrete, or continuous. To achieve the goal of statistical network models--- to understand the dependence of the edges using the observed and unobserved structure---we define the observed pair-specific covariates as $X = (x_{ij})_{(i,j) \in \mathcal{N}}$ and the unobserved latent variables as $Z= (z_{i})_{i \in \mathcal{N}}$. Specifically, $z_i$ represents an $R$-dimensional latent position (or vector) of node $i$ in the latent space models (Section \ref{sec:LSM}), and $z_i$ represents the class of node $i$ in the stochastic blockmodels where each $z_i$ takes one value in the set of categories $\mathcal{C}$ labeled as $1,\ldots, C$ (Section \ref{sec:SBM}). When extending to the dynamic models for discrete time points $t=1,\ldots,T$, we use a $t$ subscript on the variables (e.g., $y_{ijt}, x_{ijt}, z_{it}$) to denote ``at time $t$". Generally, random variables are denoted by upper case letters and fixed quantities or realizations of random variables are denoted by lower case letters. Other than the ones specified here, additional notations specific to each model are introduced later.

\section{Latent Space Models} \label{sec:LSM}
 In this section, we first describe the original latent space model introduced by Hoff et al. \cite{hoff2002latent}. Then we introduce two lines of research: (i) the latent position model \cite{handcock2007model}, which is built upon the Euclidean distance space, and (ii) the latent factor model \cite{hoff2005bilinear}, which stems from the projection model.  We present the dynamic extension of these static models and demonstrate how the two different frameworks have evolved in the dynamic network literature, in chronological order.

\subsection{Static Latent Space Models}

\subsubsection{Latent Space Approaches to Social Network Analysis}
The idea of social space is first introduced to network modeling by Hoff et al. \cite{hoff2002latent}, with the intuition that each node $i\in \mathcal{N}$ can be seen as a point $z_i$ in an $R$-dimensional space that represents unobserved latent characteristics. The presence or absence of an edge between two nodes is independent of all other edges given the latent positions of the two nodes; thus the conditional probability of the adjacency matrix $Y$ is
\begin{equation*}
    P(Y|Z, X, \theta) = \prod_{i\neq j} P(y_{ij} |z_i, z_j, x_{ij}, \theta),
    \end{equation*}
where $X$ are observed pair-specific covariates, $\theta = (\alpha, \beta)$ are the set of regression parameters. Here, the latent positions $Z$ can be interpreted as the random effects in linear models. \\

There exist two different ways to model $P(y_{ij} |z_i, z_j, x_{ij}, \theta)$: the distance model and the projection model. The main assumption here is that nodes are positioned in a latent space, and they tend to create edges to others with shorter distance (distance model) or narrower angle (projection model) between their latent positions. Both distance and projection models can be applied to various types of edges (e.g., binary, discrete, continuous) via the linked mean parameter of a generalized linear model. Without loss of generality, we introduce the model formulations for the case of binary networks.\\

\noindent\textbf{Latent Distance Model}\\

For a binary network, the distance model takes the form of a logistic regression model with $\theta = (\alpha, \beta)$ and defines the log odds of an edge between nodes $i$ and $j$ as:
\begin{equation*}
    \begin{aligned}
    \eta_{ij}&=\mbox{logodds}(y_{ij}=1|z_i, z_j, x_{ij}, \alpha, \beta)\\&
    =\alpha+\beta^\prime x_{ij} - ||z_i-z_j||,
    \end{aligned}
\end{equation*}
where $\alpha$ is an intercept term, $\beta$ is a vector of coefficients for covariate effects, and $||z_i- z_j||$ is the Euclidean distance between nodes in the latent space. In other words, the existence of an edge in the adjacency matrix (i.e., $Y_{ij} = 1$) is determined
by the dyad attributes $x_{ij}$ and the distance between the pair of nodes $ ||z_i-z_j||$. \\

Krivitsky et al. \cite{krivitsky2009representing} proposes the latent cluster random effects model that includes the distance between latent space positions, model-based clustering of the latent positions, and random sender and receiver effects (discussed below), and introduces Bayesian estimation methods for both binary and non-binary network data.

\noindent\textbf{Latent Projection Model}\\

The projection model posits a different assumption:
\begin{equation*}
    \begin{aligned}
        \eta_{ij}&=\mbox{logodds}(y_{ij}=1|z_i, z_j, x_{ij}, \alpha, \beta)\\&=\alpha+\beta^\prime x_{ij}+\frac{z_i^\prime z_j}{||z_j||},
            \end{aligned}
        \end{equation*}
which changes the interpretation to such that $i$ and $j$ are more likely to form an edge if $z_i$ and $z_j$ are in the same direction (i.e.,\ $z_i^\prime z_j>0$), while less likely to form an edge if they are in opposite directions (i.e.,\ $z_i^\prime z_j<0$). Due to the denominator term $||z_j||$, this also allows asymmetric edge probabilities (i.e., $\eta_{ij} \neq \eta_{ji}$) even when the dyadic covariates $x_{ij}$ are symmetric.

\subsubsection{Latent Position Cluster Model}
As identifying groups of similar nodes is often of interest to network researchers, Handcock et al. \cite{handcock2007model}  extends the latent space models to allow clustering of nodes in networks. Referred to as the ``latent position cluster model", this approach combines the latent space models and the model-based clustering method \cite{mbc93}.  The modeling framework is built upon the latent distance model
\begin{equation*}
\mbox{logodds}(y_{ij}=1|z_i, z_j, x_{ij}, \alpha, \beta)=\beta_0' x_{ij} - \beta_1||z_i-z_j||,
\end{equation*}
where each position $z_i \in \mathbb{R}^K$ is drawn from a finite mixture of multivariate normal distributions from $G$ groups. Assuming different means and covariance matrices for each group $g=1,...,G$, the model proposes
\begin{equation*}z_i \sim \sum_{g=1}^G \lambda_g N_d(\mu_g, \sigma_g^2I_R),
\end{equation*}
where $\lambda_g$ is the probability that an actor belongs to $g$, and $I_R$ is the $R\times R$ identity matrix.\\

Inference of the latent space model and the clustering model can be made with either the maximum likelihood method (MLE) or the Bayesian approach via Monte
Carlo Markov chain (MCMC), and those methods are implemented in the R package `latentnet' \cite{latentnet2,latentnet}.

\subsubsection{Bilinear Mixed-effects Model} \label{sec:lfm}
Providing some computational and conceptual advantages to the latent space models, Hoff \cite{hoff2005bilinear} develops the generalized bilinear mixed-effects models (GBME), a network regression model that is in line with the latent space models, especially the latent projection models. Starting from the traditional regression model on dyadic data
\begin{equation*}
y_{ij} = {\beta}^\prime {x}_{ij}+\epsilon_{ij},
\end{equation*}
this new approach includes additive and multiplicative random effects in the error terms, i.e.,
\begin{equation*}
    \begin{aligned}
    \epsilon_{ij} = a_i &+ b_j + {z}_i^\prime {z}_j+ \gamma_{ij}\\
(a_i, b_j)^\prime &\sim N({0}, \Sigma_{ab}),\\
z_i &\sim N({0}, \sigma_z^2I_R),\\
(\gamma_{ij}, \gamma_{j, i})^\prime &\sim N({0}, \Sigma_{\gamma}), \mbox{ where }\\
\Sigma_{ab} = \begin{pmatrix}
    \sigma_a^2 & \sigma_{ab} \\
    \sigma_{ab} & \sigma_b^2
\end{pmatrix}& \mbox{ and } \Sigma_{\gamma} = \sigma_\gamma^2\begin{pmatrix}
1 & \rho \\
\rho &1
\end{pmatrix}.
\end{aligned}
    \label{eqn:gbme}
\end{equation*}
This representation allows us to explain the second-order dependence often seen in dyadic data: a sender effect $a_i$, a receiver effect $b_j$, and a within dyad effect $\gamma_{ij}$. Variance parameters $\sigma_a^2$ and $\sigma_b^2$ represents the dependence of observations having a common sender and receiver, respectively, with the correlation between sender and receiver effects described by $\sigma_{ab}$. Moreover, reciprocity and within-actor correlation are measured by $\sigma_\gamma^2$ and the correlation parameter $\rho$. Finally, the bilinear effect in the product term ${z}_i^\prime {z}_j$ is known to capture the third-order dependence in dyadic data such as transitivity (i.e., if $y_{ij} = 1$ and $y_{jk} = 1$, then we tend to have $y_{ik} = 1$), balance (i.e., $y_{ij}\times y_{jk}\times y_{ki} > 0$ for signed unordered edges), and clusterability (i.e., a triad can be divided into groups).
\\

    The GBME framework has been further studied and modified in \cite{hoff2008modeling,hoff2009multiplicative,minhas2016inferential}, and has been referred to as the ``latent factor models", the ``eigenmodel'',  and the ``additive and multiplicative effects (AME) models" \cite{hoff2013likelihoods,hoff2015dyadic,minhas2016inferential}. Throughout this paper, we use the name ``latent factor models" to emphasize the objective of the paper in understanding latent variable models. \cite{hoff2008modeling} shows that the latent factor model generalizes the latent distance model and the latent class model. For the effective use of the latent factor models, the R package ``AMEN" \cite{hoff2015dyadic} provides estimation and inference for a class of AME models for ordinal, continuous, binary and other types of dyadic data.

\subsection{Dynamic Latent Distance Models}
In this section, we summarize some dynamic latent distance models which extend the latent distance models with Markovian properties.

\subsubsection{Dynamic Social Network in Latent Space Model}\label{subsubsec:DSNL}
The first dynamic latent distance model appeared in the machine learning literature by Sakar and Moore \cite{sarkar2005dynamic}. They named it the ``dynamic social network in latent space (DSNL) model''.  This model assumes that $Y_t$ (i.e., the observed pairwise adjacency matrix at time $t$) is only dependent on the current latent positions $Z_t$ (``observational model") while allowing the latent positions to move over time with standard Markovian assumptions (``transition model").\\

 The observation model for the graph can be written as
 \begin{equation*}
 P(Y_t|Z_t) = \prod_{i\sim j} p_{ijt} \prod_{i \nsim j}(1-p_{ijt}),
 \end{equation*}
 where $i\sim j$ and $i\nsim j$ denote the existence and absence of a link, respectively, and $p_{ijt}$ is the probability of a link between $i$ and $j$ at time $t$. Specifically, it has the following form:
 \begin{equation*}
 p_{ijt}=\frac{1}{1+e^{(d_{ijt}-r_{ijt})}}K(d_{ijt})+\rho (1-K(d_{ijt})),
 \end{equation*}
 where $d_{ijt} = ||z_{it}- z_{jt}||$ is the Euclidean distance between $i$ and $j$ in the latent space as in the original latent distance model at time $t$, $r_{ijt}$ is the maximum of the radii of $i$ and $j$ at time $t$, defined as $r_{ijt}\propto (max(\delta_{it}, \delta_{jt}) + 1)$ with $\delta_{it}$ being the degree of node $i$ at time $t$, $K$ is a biquadratic kernel $K(d_{ijt})=(1-(d_{ijt}/r_{ijt})^2)^2$, and $\rho$ is a constant noise. \\

 The transition model assumes a Gaussian random walk:
 \begin{equation*}
 \begin{aligned}
 Z_t|Z_{t-1} &\sim N(Z_{t-1}, \sigma^2I),
 \end{aligned}
 \end{equation*}
 and the resulting log$P(Z_t|Z_{t-1})$ becomes
 \begin{equation*}
 \begin{aligned}
 \mbox{log}P(Z_t|Z_{t-1})&\propto-\sum_{i=1}^n |z_{it}-z_{i(t-1)}|^2/2\sigma^2,
 \end{aligned}
 \end{equation*}
 such that we want to estimate the positions using
\begin{equation*}
    \begin{aligned}
&P(Z_t |Y_t, Z_{t-1}) \propto P(Y_t|Z_t) P(Z_t|Z_{t-1}).\\
\end{aligned}
\end{equation*}

For efficient optimization of the likelihood, the authors develop the two-stage learning algorithms: (i) generalized multidimensional scaling (MDS) to find the initial latent positions across time and (ii) nonlinear conjugate gradient (CG) optimization starting from the initial estimates. Additionally, the Procrustean transformation is applied to the coordinates from MDS so that $Z_t$ maintains the same orientation as $Z_{t-1}$. \\

The DSNL model is applied to NIPS conference paper co-authorship dataset (network size up to 11,000) by separating the 12 years of data into three discrete time points: 1987--1990, 1991--1994, and 1995--1998. The analysis is mainly focused on investigating the dynamics of some well-connected researchers in the machine learning community, and the authors find out some noticeable changes in the network embeddings via the examination of latent positions. Sakar et al. \cite{sarkar2007latent} further extends the model to be applicable for dynamic bipartite or two-mode networks, such as author-word data from the same NIPS data using text corpora.

\subsubsection{Dynamic Latent Distance Model with Popularity and Activity Effects}
Sewell and Chen \cite{sewell2015latent} propose a network model for longitudinal networks which allows each node to have a temporal trajectory in an $R$-dimensional latent Euclidean space, with additional features to capture the nodes' popularity and activity effects.\\

For a binary network $Y_t$, the model formulation relies on the log-odds $\eta_{ij}$:
\begin{equation*}
    \begin{aligned}
P({Y}_t|{Z}_t, \beta_{IN},\beta_{OUT},r_{1:N}) &= \prod_{i\neq j} \frac{\mbox{exp}(y_{ijt}\eta_{ijt})}{1+\mbox{exp}(y_{ijt}\eta_{ijt})}, \mbox{ where }\\
\eta_{ijt}&=\beta_{IN}(1-\frac{d_{ijt}}{r_j})+\beta_{OUT}(1-\frac{d_{ijt}}{r_i}),
\end{aligned}
\end{equation*}
where the distance between $i$ and $j$ at time $t$ is defined as $d_{ijt} =||{z}_{it}-{z}_{jt}||$ and $r_i$ is a vector of positive actor-specific parameters with constraints $\sum_{i=1}^N r_i = 1$. Specifically, $r_i$ can be thought of as the ith actor's social reach or radius which does not vary over time. In addition, this formulation separates the two parameters $\beta_{IN}$ and $\beta_{OUT}$ to measure the global popularity and activity effects, respectively. Note that this model deals with directed networks (i.e., $y_{ij} \neq y_{ji}$) as well as undirected ones (i.e., $y_{ij} =y_{ji}$), while the DSNL model in Section \ref{subsubsec:DSNL} only allows undirected edges.\\

Similar to the DSNL model, the latent positions ${Z}_t$ at time $t$ are modeled by a Markov process with a transition equation for $t=2,3,...,T$:
\begin{equation*}
\pi({Z}_t|{Z}_{t-1}, \phi) = \prod_{i=1}^n N({z}_{it}|{z}_{i(t-1)}, \sigma^2I_R),
\end{equation*}
where the initial positions have independent ${z}_{i1}\sim N({0}, \tau^2I_R)$ for $i=1,\ldots,N$ and $\phi=(\tau^2,\sigma^2,\beta_{IN},\beta_{OUT},r_{1:n})$ is the set of parameters of interest. To estimate the parameters $\phi$, the authors adopt a Bayesan approach and implement MCMC algorithm using Metropolis-Hastings within Gibbs. They also perform a Procrustes transformation in order to re-orient the sampled trajectories. \\

This dynamic latent distance model is applied to two datasets. The Dutch classroom dataset includes directed relationships among 25 students, surveyed over four time points.  The bill cosponsorship dataset consists of the cosponsorship history of 644 members of Congress who served during the 97th to 101st Congresses in the U.S. House of Representatives (five time points). The model provides useful insights into the networks, such as the overall gender and ethnic separation in the Dutch classroom example and the reflection of the political ideology of the latent positions in the bill cosponsorship example. In their follow-up work, Sewell and Chen \cite{sewell2016latent} adjust the model for weighted edges including rank-ordered count data and non-negative continuous edges, via link functions and data augmentation. The applicability and interpretability of this extended model are demonstrated with mobile phone call log data and world exports/imports data, which are the examples of the count and continuous edges, respectively.

\subsubsection{Dynamic Latent Distance Model for Bipartite Network}
A bipartite graph is a graph in which vertices are divided into two groups and links only exist across groups. Friel et al. \cite{friel2016interlocking} develop a statistical model for bipartite temporal networks by extending the DSNL model, with different assumptions from the bipartite network model in Sakar et al. \cite{sarkar2007latent}. While Sakar et al. \cite{sarkar2007latent} model the co-occurrence of counts of authors and words from their empirical distribution, Friel et al. \cite{friel2016interlocking} directly model the evolution of edges through three Markov processes: one on the parameters, one on the latent positions, and one on the edges.\\

Define the binary edge $y_{ijt}=1$ if node $i$ in the first group ($i = 1,\ldots, N$) is connected to node $j$ in the second group ($j = 1,\ldots, M$) at time $t$ (e.g., director $i$ sitting on board $j$ in year $t$), and $y_{ijt}=0$ otherwise. The chance of forming an edge at time $t$ depends on the previous state $y_{ij(t-1)}$, and the latent distance between the two nodes in two different groups. The model assumes that
\begin{equation*}
\mbox{logit}(p_{ijt}) = y_{ij(t-1)}\gamma_t + (1-y_{ij(t-1)})\beta_t-||{z}_{it}-{w}_{jt}||,
\end{equation*}
where $\gamma_t$ is an edge persistence parameter and $\beta_t$ is a non-edge persistence parameter, which are separately defined to capture the difference in the persistence of edges and non-edges, and $z_{it}$\footnote{While the general framework assumes time-varying latent positions $z_{it}$, Friel et al. \cite{friel2016interlocking} fixed the first group's latent positions (i.e., $z_i$) for their application.} and $ w_{jt}$ are the $R$-dimensional latent positions of node $i$ in the first group and $j$ in the second group, respectively, at time $t$. \\

The latent positions in the first group are assumed to have independent Gaussian prior, while the latent positions in the second group are modeled using random walk processes (given the initial positions at $t=1$):
\begin{equation*}
\begin{aligned}
P(Z) &= \prod\limits_{t=1}^T\prod\limits_{i=1}^N\prod\limits_{r=1}^R f(z_{irt}; {0}, \tau_z),\\
P(W|\tau_w, \tau_w^0) &= \prod\limits_{j=1}^M\prod\limits_{r=1}^R f(w_{jr1}; {0}, \tau_w^0)\times\prod\limits_{t=2}^T\prod\limits_{j=1}^M\prod\limits_{r=1}^R f(w_{jrt}; w_{jr(t-1)}, \tau_w),\\
\end{aligned}
\end{equation*}
where $(\tau_w, \tau_w^0, \tau_z)$ are the precision parameters. Same as the latent positions of the second group, this model further assumes that the parameters $\gamma_t$ and $\beta_t$ also follow the Markov processes:
\begin{equation*}
    \begin{aligned}
        P({\gamma}|\tau_\gamma, \tau_\gamma^0) &= f(\gamma_1; {0}, \tau_\gamma^0)\times\prod\limits_{t=2}^Tf(\gamma_t; \gamma_{t-1}, \tau_\gamma),\\
    P({\beta}|\tau_\beta, \tau_\beta^0) &= f(\beta_; {0}, \tau_\beta^0)\times\prod\limits_{t=2}^Tf(\beta_t; \beta_{t-1}, \tau_\beta),
    \end{aligned}
\end{equation*}
where $(\tau_\gamma, \tau_\gamma^0, \tau_\beta, \tau_\beta^0)$ are the precision parameters. Same as the inference procedures in other dynamic latent distance models, MCMC sampling is used to sample from the joint posterior distribution, and Procrustes transformation is applied for fixed orientation.
\\

Using the model, the authors analyze the dynamic evolution of the leading Irish companies and their directors from 2003 to 2013. The network data contain $N=761$ directors (first group), $M=59$ companies (second group), and 3,855 edges. Mainly focused on understanding the persistence of links and the heterogeneity in the latent positions, the analysis reveals an increasing level of interlocking board behavior before and during the financial crisis, and stabilization thereafter.

\subsection{Dynamic Latent Factor Models}

As mentioned before, the multiplicative form in the projection model \cite{hoff2002latent} has been continuously studied under the name of ``latent factor models". As shown in Hoff \cite{hoff2008modeling}, latent factor models have some advantages over the latent distance model when both homophily---a type of pattern in which similar nodes may be more likely to attach to each other than dissimilar ones---and stochastic equivalence---a type of pattern in which the nodes can be divided into groups such that members of the same group have similar patterns of relationships---are present in the network. In this section, we demonstrate the development of dynamic latent factor models along with the motivating examples that highlight the models' capability of capturing higher-order dependence such as reciprocity---tendency to form respective edges in a directed network---and stochastic equivalence.

\subsubsection{Dynamic Gravity Model}
Gravity models \cite{leibenstein1966shaping} in social sciences are used to model the bilateral relationships that can be predicted by the mass and distance of the the pair.
In modeling international trade and conflicts, Ward and Hoff \cite{ward2007persistent} and Ward et al. \cite{ward2007disputes} are the earlier works that combine the bilinear mixed-effects model (GBME) in Section \ref{sec:lfm} with the gravity models. However, their applications to temporal networks are done by separately fitting the static model to each time point's network. Later, Ward et al. \cite{ward2013gravity} improve their approach by including the previous year's fitted positions. The resulting model has the following form:
\begin{equation*}
\begin{aligned}
\mbox{log}(y_{ijt}) & = \beta'x_{ijt}+\beta_{uv}{\hat u}'_{i(t-1)}{\hat v}_{j(t-1)}+a_{it}+b_{jt}+{u}'_{it}{v}_{jt}+\epsilon_{ij},
\end{aligned}
\end{equation*}
where $x_{ijt}$ consists of three covariates in the gravity models---the log of gross domestic product (GDP) of country $i$ and $j$ at time $t$, respectively, and the log of geographic distance between country $i$ and $j$ at time $t$. Different from the GBME, this model separates the latent positions to sending activity $u$ and receiving activity $v$, and also includes the lagged position term $\beta_{uv}{\hat u}'_{i(t-1)}{\hat v}_{j(t-1)}$ to estimate the effect of positions at time $t-1$ on the edges at time $t$.\\

This model is applied to bilateral trade data from 1990 to 2008. After several goodness of fit tests including the $R^2$ checks and out-of-sample predictions, the authors confirm the presence of strong second-order (i.e., reciprocity) and third-order dependencies in the world trade network. They demonstrate that the GBME model outperforms conventional ones in international trade literature, regarding predicting the observed values and understanding unobserved dyadic relationships.

\subsubsection{Hierarchical Multilinear Model}

By treating dynamic network data as an example, Hoff \cite{hoff2011hierarchical,hoff2011separable} develops a general modeling framework for array data via reduced-rank decompositions, in which the array can be expressed as products of low-dimensional latent factors. Specifically for symmetric dynamic network data $Y_t$, the model takes the following form:
\begin{eqnarray*}
Y_t &=& X_t \beta +  \Gamma_t,\\
\Gamma_t &=&  Z \Lambda_t  Z' +  E_t,
\end{eqnarray*}
where $Z=(z_1, ...,z_N)'$ is a matrix consisting of the time-invariant eigenvectors, assumed to have a matrix normal prior distribution, and $\Lambda_t=\mbox{diag}(\lambda_{1t},\ldots,\lambda_{Rt})$ is the time-varying eigenvalue matrix. This can be seen as a generalization of the latent factor models, using the reduced-rank approximations to matrices. Further extensions to arrays can be found in Hoff (2015b) \cite{hoff2015multilinear}, with longitudinal network serving as an example of the general model. This model is implemented as the R package ``AMEN" \cite{hoff2015dyadic}.\\

In Hoff \cite{hoff2011separable}, the model is applied to analyze the international cooperation and conflict data in Ward et al. \cite{ward2007disputes}. The dataset includes the records of militarized conflict and cooperation of 66 countries in every five years from 1950 to 1985, along with some economic and political characteristics of the countries. The model discovers that a majority of the conflicts over the cold war period involves economically large countries, and also reveals different conflict and cooperation patterns across countries over time.

\subsubsection{Dynamic Bilinear Effects Model}\label{subsubsec:Dunson}
Gaussian processes (GP's) have long been used in modeling temporally or spatially dependent data. Durante and Dunson \cite{durante2014bayesian,durante2014} propose a model that extends the bilinear component in Section \ref{sec:lfm} to dynamic case by assuming the mean parameters and the latent factors are evolving in time via Gaussian processes. In particular, the model assumes
\begin{eqnarray*}
\eta_{ijt}&=&\mu_t+\beta'_tx_{ijt}  + z_{it}'z_{jt},
\end{eqnarray*}
where $\eta$ is the linked mean parameter of a generalized linear model, $\mu_t$ is the mean process, and $z_{it}$ and $z_{jt}$ are the $R$-dimensional latent positions at time $t$. Here, the parameters $\mu$, $\beta$, and $z_{ir}$ (for $i = 1,\ldots, n, \mbox{ and } r = 1,\ldots, k$) vary over time given their GP priors
\begin{eqnarray*}
(\mu_1,\ldots,\mu_T)' &\sim& N(0,c_{\mu}),\\
(z_{ir1},\ldots,z_{irT})' &\sim& N(0,\tau_r^{-1}c_z),
\end{eqnarray*}
where $c_\mu$ and $c_z$ are both $T\times T$ matrices constructed using squared exponential covariance function---as a function of the distance between two timepoints $t_1$ and $t_2$. In other words,
\begin{eqnarray*}
c_{\mu}(t_1,t_2)&=&\exp\{-\kappa_{\mu}(t_1-t_2)^2\},\\
c_z (t_1,t_2)&=&\exp\{-\kappa_z(t_1-t_2)^2\}, \\
\tau_r = \prod_{l=1}^r \upsilon_l, \quad \upsilon_1 &\sim& \mbox{Ga}(a_1,1), \quad \upsilon_l \sim \mbox{Ga}(a_2,1), \quad l \ge 2,
\end{eqnarray*}
where $\kappa_\mu$ and $\kappa_z$ are the length-scale parameters in GP, and $\tau_r$ is the variance parameter with hierarchical Gamma priors.
The main advantage of this model is that it relaxes the Markovian assumption in most of the previously mentioned work and allows an unequal spacing of the observed time points.\\

 Durante and Dunson \cite{durante2014} apply the method to analyze the co-movements of 23 National Stock Market Indices from 2004 to 2013. The results successfully reflect the global financial crisis and Greek debt crisis periods. The time-varying predictors help to discover the existence of international financial contagion effects and the opposite effects of verbal and material cooperation efforts on financial co-movements.

\section{Stochastic Blockmodels} \label{sec:SBM}

In this section, we briefly summarize a statistical approach to \emph{a} \emph{posteriori} block modeling for networks introduced by \cite{snijders-1997} and \cite{nowicki-2001} where the membership of nodes are unknown as well as the two relevant models, the mixed membership stochastic blockmodel \cite{airoldi2008mixed} and the degree-corrected stochastic blockmodel \cite{karrer-newman-2011}, and then introduce several dynamic extensions of these static models.


\subsection{Static Stochastic Blockmodels}

The main assumption of the stochastic blockmodels is that the nodes of the network are partitioned into several latent classes (or blocks). To be specific, Snijders and Nowicki \cite{snijders-1997} and Nowicki and Snijders \cite{nowicki-2001} assume the set of nodes $i \in \mathcal{N}$ is partitioned into $C$ categories labeled $1, \ldots, C$. The class of node $i$ is denoted by $z_{i}$ and the classes are obtained in the vector ${z} = (z_{i})_{i=1}^{N}$. The set of classes is denoted by $\mathcal{C} = \{1, \ldots,C\}$. The models further assume that the probability distribution of the edge between two nodes depends only on the classes to which they belong. Thus the edges are conditionally independent given the class vector. \\

Stochastic blockmodels inherit the philosophy of finite mixture models, and assume that the unobserved classes $Z_{i}$ are \emph{i.i.d.} random variables with the probability
\begin{equation*}
    P(Z_{i} = c) = \theta_{c},
\end{equation*}
for class $c \in \mathcal{C}$. Therefore, the joint distribution of ${Z} = \{Z_{1},...,Z_{N}\}$ is defined by
\begin{equation*}
    P(Z_{1} = z_{1}, \ldots, Z_{N}=z_{N}) = \theta_{1}^{m_{1}} \cdots \theta_{C}^{m_{C}},
\end{equation*}
where
\begin{equation*}
    m_{c} = \sum_{i=1}^{N}I(z_{i} = c)
\end{equation*}
denotes the number of vertices with class $c$. \\

The model for the edges between the nodes depends on the node classes in the following ways. Given the vector of node classes ${Z} = {z}$, the random vectors ${Y}_{ij}$ for $(i,j) \in \mathcal{N}$ with $i<j$ are independent, and the probabilities are
\begin{equation*}
    P({Y}_{ij} = {y} \mid {Z} = {z}) = \eta_{{y}}(z_{i},z_{j}),
\end{equation*}
where ${y} = (y_{i\rightarrow j}, y_{j \rightarrow i}) \in \mathcal{Y}$ is a vector of edge values and $\eta_{{y}}(z_{i},z_{j})$ is the class dependent edge probabilities which satisfy the condition
\begin{equation*}
    \sum_{{y} \in \mathcal{Y}} \eta_{{y}}(g,h) = 1 \quad \mbox{for all} \quad g, h \in \mathcal{C}.
\end{equation*} \newline
The conditional distribution of ${Y}$ given the vector of classes ${z}$ is given in the following form
\begin{equation*}
    P({Y} = {y} \mid {z}, {\theta}, {\eta}) = \left(\prod_{{y} \in \mathcal{Y}} \prod_{1\leq g < h \leq C}(\eta_{{y}}(g, h))^{e_{{y}}(g,h)}\right) \times \left(\prod_{{y} \in \mathcal{Y}} \prod_{c=1}^{C}(\eta_{{y}}(c, c))^{e_{{y}}(c,c)}\right),
\end{equation*}
where $e_{{y}}(g,h)$, ${y} \in \mathcal{Y}$, $1 \leq g \leq h \leq C$, are the edge counts for block $(g,h)$. \\

For the inference of the stochastic blockmodels, both Bayesian and frequentist approaches are proposed. In Bayesian approach, the prior distribution for parameters is taken to be a product of Dirichlet distributions for the class distribution and the edges between and within classes of the given memberships. The posterior distribution is estimated using the Gibbs sampler. On the other hand, in frequentist approach, several different algorithms are proposed to estimate the classes: Rohe et al. \cite{rohe-etal-2011} propose the spectral clustering to estimate classes in the model, Gu\'{e}don and Vershynin \cite{guedon-vershynin-2016} and Amini and Levina \cite{amini-levina-2018} write the class estimation problem as semidefinite optimization problem and find the solution, and Amini et al. \cite{amini-etal-2013} propose pseudo-likelihood algorithm which provides consistent estimates of classes. \\


Despite the simplicity in model formulation, stochastic blockmodels provide a powerful tool in modeling networks. They allow one to represent the effect of unobserved heterogeneity of individual positions or preferences on the pattern of pairwise relations. Since the heterogeneity is modeled by the stochastic membership of the $c$ classes, regarding cluster analysis, it is more similar to a mixture model rather than a discrete classification model.

\subsubsection{Mixed Membership Stochastic Blockmodel}

Many real-world networks are multi-faceted. However, stochastic blockmodels suffer from a limitation that each node can only belong to one group, or in other words, play a single latent role. To overcome this issue, Airoldi et al. \cite{airoldi2008mixed} relax the assumption of a single latent role for nodes and develop the mixed membership stochastic blockmodel. \\

In this paper \cite{airoldi2008mixed}, the authors focus on directed networks and assume the observed network is generated according to node-specific distributions of community membership and edge-specific indicator vectors denoting the membership in one of the $C$ communities. Each node is associated with a randomly drawn vector $\vec{\pi}_{i}$ for node $i$, where $\pi_{i,c}$ denotes the probability of node $i$ belonging to group $c$. That is, each node can simultaneously belong to multiple groups with different degrees of affiliation degree. The probabilities of edges between different groups are defined by the matrix of Bernoulli rates ${B}_{C\times C}$, where ${B}(g, h)$ represents the probability of having an edge between a node from group $g$ and a node form group $h$. The mixed membership stochastic blockmodel posits that the $\{Y_{ij}\}_{1<i,j<N}$ are drawn from the following generative process. \\
\begin{itemize}
    \item For each node $i=1,\ldots,N$:
    \begin{itemize}
        \item Draw a $C$ dimensional mixed membership vector $\vec{\pi}_{i} \sim \mbox{Dirichlet}(\vec{\alpha})$.
    \end{itemize}
    \item For each possible edge variable $Y_{ij}$:
    \begin{itemize}
        \item Draw membership indicator for the initiator $\vec{z}_{i \rightarrow j} \sim \mbox{Multinomial}(\vec{\pi}_{i})$.
        \item Draw membership indicator for the receiver $\vec{z}_{i \leftarrow j} \sim \mbox{Multinomial}(\vec{\pi}_{j})$.
        \item Sample the edge $Y_{ij} \sim \mbox{Bernoulli}(\vec{z}'_{i \rightarrow j}B\vec{z}_{i \leftarrow j})$.
    \end{itemize}
\end{itemize}
Here, the indicator vector $\vec{z}_{i \rightarrow j}$ denotes the group membership of node $i$ when node $i$ has out-going edge to node $j$, and $\vec{z}_{j \rightarrow i}$ denotes the group membership of node $j$ when node $j$ has out-going edge to node $i$. \\

Under the mixed membership stochastic blockmodel assumptions, the joint probability of the data $Y$ and the latent variables $\{\vec{\pi}_{1:N},Z_{\rightarrow},Z_{\leftarrow}\}$ can be written in the following form:
\begin{equation*}
    \begin{split}
        &P(Y,\vec{\pi}_{1:N},Z_{\rightarrow},Z_{\leftarrow} \mid \vec{\alpha}, B) \\&= \prod_{(i,j) \in \mathcal{N}}P(Y_{ij} \mid \vec{z}_{i \rightarrow j}, \vec{z}_{i \leftarrow j}, B) P(\vec{z}_{i \rightarrow j} \mid \vec{\pi}_{i}) P(\vec{z}_{i \leftarrow j} \mid \vec{\pi}_{j})
\times \prod_{i=1}^NP(\vec{\pi}_{i} \mid \vec{\alpha})
    \end{split}
\end{equation*}
where $\vec{\pi}_{1:N}$ is the set of mixed membership vectors, and $Z_{\rightarrow}$ and $Z_{\leftarrow}$ are the sets of membership indicator vectors. \\

A nested variational inference algorithm is used for posterior inference on the per-node mixed membership vectors and per-pair roles. To compute the empirical Bayes estimates of the model parameters, variational expectation-maximization (EM) algorithm is used. Moreover, in recent years, Mao et al. \cite{mao-etal-2017} and Jin et al. \cite{jin-etal-2017} propose consistent algorithm inferring mixed membership of nodes in the mixed membership stochastic blockmodel.

\subsubsection{Degree-Corrected Stochastic Blockmodel}

Although the stochastic blockmodels have been popularly used as a tool for detecting community structure in networks, they fail to capture the heterogeneity of node degrees (i.e., number of edges the node has to other nodes) within communities which often observed in real-world networks. To solve this problem, Karrer and Newman \cite{karrer-newman-2011} relax the assumption that the stochastic blockmodels treat all nodes within a community as stochastically equivalent, and propose the degree-corrected stochastic blockmodel that can consider node covariates. \\

This model focuses on undirected networks and allows networks to contain both multi-edges and self-edges, even though many real-world networks have no such edges. Let $Y$ be an undirected multigraph on $n$ nodes, possibly including self-edges, and let $Y_{ij}$ be an element of the adjacency matrix of the multigraph. Also, there is a new set of parameters $\theta_{i}$ controlling the expected degrees of nodes $i$. The model assumes that the number of edges between each pair of nodes is independently Poisson distributed, and define $\theta_{i}\theta_{j}\omega_{gh}$ to be the expected value of the adjacency matrix element $Y_{ij}$ for nodes $i$ and $j$ lying in groups $g \in \mathcal{C}$ and $h \in \mathcal{C}$, respectively. Then, $Y$ has the probability
\begin{equation*}
    \begin{split}
        P(Y \mid \theta, \omega, z) =& \prod_{i<j}\frac{(\theta_{i}\theta_{j}\omega_{z_{i}z_{j}})^{Y_{ij}}}{Y_{ij}!}\exp(-\theta_{i}\theta_{j}\omega_{z_{i}z_{j}}) \\
        & \times \prod_{i}\frac{(\frac{1}{2}\theta_{i}^{2}\omega_{z_{i}z_{j}})^{Y_{ij}/2}}{(Y_{ij}/2)!}\exp(-\frac{1}{2}\theta_{i}^{2}\omega_{z_{i}z_{j}}).
    \end{split}
\end{equation*} \newline

To estimate the parameters and infer the membership of nodes, the authors suggest a novel method where the log-likelihood is maximized in two stages. First, they find maximum likelihood values of model parameters $\theta_{i}$ and $\omega_{gh}$. Then, they propose information-theoretic quantities for community detection or clustering. Similar to Rohe et al. \cite{rohe-etal-2011}, Qin and Rohe \cite{qin-rohe-2013} propose spectral clustering under the degree-corrected stochastic blockmodel. Also, Zhao et al. \cite{zhao-etal-2012} generalize the consistency framework of stochastic blockmodel to the degree-corrected stochastic blockmodel and obtain a general theorem for community detection consistency.

\subsection{Dynamic Stochastic Blockmodels}

\subsubsection{Dynamic Stochastic Blockmodel with Varying Community Memberships}
To analyze dynamic communities, Yang et al. \cite{yang-etal-2011} propose a model that captures the evolution of communities by explicitly modeling the transition of community memberships for individual nodes in the network. \\

Let $Y_t$ be the snapshot of a network at a given time step $t$, with the $N$ number of nodes in the network. Each element $y_{ijt}$ in $Y_t$ is the weight assigned to the edge between nodes $i$ and $j$. Although this dynamic stochastic blockmodel can handle both the frequency of interactions (i.e., a natural number) and the binary number indicating presence or absence of interactions, the authors' main focus is on the binary edges. For a dynamic network, they use $\mathcal{Y}_{T} =\{Y_1, \ldots, Y_T\}$ to denote a collection of snapshots for the network over $T$ discrete time steps. They also use $z_{i} \in \{1, \ldots, C\}$, where $C$ is the total number of communities, to denote the community membership of node $i$. In addition, they introduce $z_{ic} = [z_{i} = c]$ to indicate if node $i$ is in the $c$th community where $[x]$ outputs 1 if $x$ is true and zero otherwise. Community matrix $Z_t = (z_{ict}: i \in \{1, \ldots, N\}, c \in \{1, \ldots, C\})$ then indicates the community assignments of all nodes in network at a given time step $t$. Finally, they set $\mathcal{Z}_{T} = \{Z_1, \ldots, Z_T\}$ to represent the collection of community assignments of all nodes over $T$ time steps. \\

Assuming that the community matrix $Z_{t-1}$ for time step $t-1$ is given, the model uses a transition matrix $A \in \mathbb{R}^{C \times C}$ to model the community matrix $Z_t$ at time step $t$. Moreover, $\pi = \{\pi_{1}, \ldots, \pi_{C}\}$ is used as initial probability where $\pi_{c}$ is the probability for a node to be assigned to community $c$. Given the community memberships in $Z_t$, the edge between nodes is determined stochastically by probabilities $P$. \\

The joint probability of the data $\mathcal{
    Y}_{T}$ and the latent variable $\mathcal{Z}_{T}$ can be written in the following form:
\begin{equation*}
    P(\mathcal{Y}_{T}, \mathcal{Z}_{T} \mid \pi, P, A) = \prod_{t=1}^{T} P(Y_t \mid Z_t, P) \prod_{t=2}^{T}P(Z_t \mid Z_{t-1}, A)P(Z_1\mid \pi)
\end{equation*}
where
\begin{equation*}
    \begin{aligned}
        P(Y_t \mid Z_t, P) = & \prod_{i \sim j}P(y_{ijt} \mid z_{it}, z_{jt}, P) \\
        = & \prod_{i \sim j}\prod_{g,h \in \mathcal{C}}\left(P_{gh}^{y_{ijt}}(1-P_{gh})^{1-y_{ijt}}\right)^{z_{igt}z_{jht}},\\
        P(Z_t\mid Z_{t-1}, A) = & \prod_{i=1}^{N}P(z_{it} \mid z_{i(t-1)}, A) \\
        = & \prod_{i=1}^{N}\prod_{g,h \in \mathcal{C}}A_{kl}^{z_{ig(t-1)}z_{iht}},
    \end{aligned}
\end{equation*}
and
\begin{equation*}
    P(Z_1 \mid \pi) =  \prod_{i=1}^{N}\prod_{c=1}^C\pi_{c}^{z_{ic1}},
\end{equation*}
respectively. Note that in their model self-loops are not considered and so in the above equations $i \sim j$ means over all $i$'s and $j$'s such that $i \neq j$. \\

For the inference of the model, the authors introduce point estimation approach using a Variational EM algorithm. They also propose a method based on a combination of probabilistic simulated annealing algorithm and Gibbs sampling algorithm to infer the parameters. \\


This model is applied to three datasets. The southern women dataset records the attendance of 18 women to 14 social events over a 9-month period in 1930's in Natchez, Mississippi, as part of their work to study social class in both black and white societies. The blog dataset is collected by the NEC labs and includes 148,681 entry-to-entry links among 407 blogs during 15 months. The paper co-authorship (DBLP) dataset contains the co-authorship information of papers in 28 conferences in three areas (data mining (DM), database (DB), and artificial intelligence (AI)) over ten years (1997-2006). For the first two datasets, the model can detect the change of community membership over the study period. In the co-authorship dataset, the model discovers the trend that, along the time, the community of DB gets smaller, the community of DM gets larger, and the community of AI remains relatively stable. Moreover, they also find some highly productive researchers had many changes in community membership.

\subsubsection{Dynamic Stochastic Blockmodel with Varying Community Memberships and Connectivity Parameters}

Both Xu and Hero \cite{xu-hero-2014} and Matias and Miele \cite{matias-miele-2017} propose methods which relax the constraint of fixed connectivity probabilities $P$ of Yang et al. \cite{yang-etal-2011} and consider both community memberships and connectivity parameters vary over time. The main difference between the two papers is that the former entirely relax the constraint of fixed connectivity parameters while the latter kept weak constraint on the connectivity parameters to handle the label switching issues across different time steps. Here we first introduce Xu and Hero \cite{xu-hero-2014} followed by Matias and Miele \cite{matias-miele-2017}. \\

Xu and Hero \cite{xu-hero-2014} propose a state space model through time on the probability of connection between groups and focus on directed networks with no self-edges. Let $Y_t$ denotes the adjacency matrix of the network observed at time step $t$. Let $\mathcal{Y}_{t}$ denotes the set of all snapshots up to time $t$, i.e., $\{Y_{1}, \ldots Y_{t}\}$. The notation $i \in c$ indicates that node $i$ belong class $c$ and the classes of all nodes at time $t$ is given by a vector $z_t$ with $z_{it} = c$ if $i \in c$ at time $t$. $\Theta_{t} = [\theta_{ght}]$ denotes the connectivity parameter between groups, where $\theta_{ght}$ denotes the probability of forming an edge between a node in class $g$ and a node in class $h$, and $C$ denotes the number of classes. The set $\Theta_t$ can be viewed as the states of a dynamic system that generates the noisy observation sequence. Since $\theta_{ght}$ is a probability that must be bounded between 0 and 1, they work with the $\psi_{ght} = \log(\theta_{ght}) - \log(1-\theta_{ght})$, which is the logit of $\theta_{ght}$. \\

The following linear dynamic system gives a model for the state evolution
\begin{equation*}
    {\psi}_{t} = F_t{\psi}_{t-1} +{v}_{t}
\end{equation*}
where $F_{t}$ is the state transition model applied to the previous state, ${\psi}_{t}$ is the vector representation of the $[\psi_{ght}] \in \mathbb{R}^{C \times C}$, and ${v}^{t}$ is a random vector of zero-mean Gaussian entries, commonly referred to as process noise, with covariance matrix $\Gamma_{t}$. They assume $\Gamma_{t}$ to be time-invariant and not necessarily diagonal because states could evolve in a correlated manner. \\

Since the class memberships $z_{t}$ are not known and should be estimated along with ${\psi}_{t}$, the label-switching methods are used as in \cite{karrer-newman-2011}, \cite{zhao-etal-2012}, and maximized the posterior state density given the entire sequence of observations $\mathcal{Y}_t$ up to time $t$, in order to account for the prior information. The posterior state density is given by
\begin{equation*}
    P({\psi}_{t} \mid \mathcal{Y}_t) \propto P(Y_t \mid {\psi}_{t}, \mathcal{Y}_{t-1}) P({\psi}_{t} \mid \mathcal{Y}_{t-1}).
\end{equation*}

Their inference is based on on-line iterative estimation procedure alternating two steps. First, a label-switching method is used to explore the space of node group configurations. Then, they use extended Kalman filter (EKF) that optimizes the likelihood when the group memberships are known. \\

This model is applied to two datasets. The MIT reality mining dataset records cell phone activity of 94 students and staff at MIT over a year. They use the participant affiliations as ground-truth class memberships and compare the class estimation accuracy of their model to static stochastic blockmodel. The Enron email dataset consists of about 500,000 email messages of 184 Enron employees from 1988 to 2002. The model discovers a steady increasing trend in edge probabilities from Enron CEOs to presidents as Enron's financial situation worsened. On the other hand, edge probabilities between other employees remained at their baseline levels until Enron fell under federal investigation. \\

Matias and Miele \cite{matias-miele-2017} focus on detecting communities characterized by a stable within-group connectivity behavior by adding some constraints on the varying connectivity parameter. They consider weighted interactions between $N$ nodes recorded over time in a set of data matrices ${Y} = (Y_{t})_{1\leq t \leq T}$. For each $t \in \{1, \ldots, T\}$, the adjacency matrix $Y_{t} = (Y_{ijt})_{1 \leq i\neq j \leq N}$ contains real-values measuring interactions between the nodes $i,j \in \{1, \ldots, N\}^{2}$. Their model can also consider undirected networks without self-loops. They assume that the $N$ nodes are split into $C$ latent communities that vary through time, as encoded by the random variables ${Z} = (Z_{it})_{1 \leq t \leq T, 1 \leq i \leq N}$. They use independent Markov chains to model the evolution of the nodes memberships over time. For each node $i$, the process $(Z_{it})_{1 \leq t \leq T}$ is an irreducible, aperiodic stationary Markov chains with transition matrix $A = (A_{gh})_{1 \leq g,h \leq C}$ and initial stationary distribution ${\alpha} = (\alpha_{1}, \ldots, \alpha_{C})$. They consider $Z_{it}$ as a random vector $Z_{it}= (Z_{i1t}, \ldots, Z_{iCt}) \in \{0,1\}^{C}$ constrained to $\sum_{c=1}^CZ_{ict} = 1$. \\

To take account of possible sparse weighted networks, they assume that
\begin{equation*}
    Y_{ijt} \mid \{Z_{igt}Z_{jht}=1\} \sim (1-\beta_{ght})\delta_{0}(\cdot) + \beta_{ght}F(\cdot, \gamma_{ght}),
\end{equation*}
where $g, h\in \mathcal{C}$ and $\{F(\cdot, \gamma), \gamma \in \Gamma\}$ is a parametric family of distribution with no mass at 0 with its density denoted by $f(\cdot, \gamma)$. Here $\beta_{t} = (\beta_{ght})_{1 \leq g,h \leq C}$ is the sparsity parameter which satisfy $\beta_{ght} \in [0, 1]$. Moreover, $\gamma_{ght}$ is the connectivity parameter depends on the choice of the parametric family. They also let $\phi(\cdot; \beta, \gamma)$ to denote the density of the distribution. \\

The joint probability of the data ${Y}$ and the latent variable ${Z}$ can then be written in the following form:
\begin{equation*}
    \begin{split}
        P({Y}, {Z} \mid {A}, {\beta}, {\gamma}) =& \prod_{i=1}^{N}\prod_{c=1}^{C}\alpha_{c}^{Z_{ic1}} \prod_{t=2}^{T}\prod_{i=1}^{N}\prod_{1 \leq g, h  \leq C}A_{gh}^{Z_{ig(t-1)}Z_{iht}} \\
        & \times \prod_{t=1}^{T}\prod_{1\leq i < j \leq N}\prod_{1 \leq g, h \leq C}\phi(Y_{ijt}; \beta_{ght}, \gamma_{ght})^{Z_{igt}Z_{jht}}.
    \end{split}
\end{equation*}

To infer the model parameters and to cluster the nodes, variational expectation-maximization (VEM) algorithm is employed. Furthermore, they propose integrated classification likelihood (ICL) criterion for estimating the number of communities. \\

This model is applied to reveal social structure and organizations in a French high school and two animal interaction datasets. The French high school dataset consists of face-to-face encounters of 31 students in the class during four days in December 2011. The model finds four distinct groups showing different interaction patterns. The model also discovers the evidence of some gender homophily. The migratory birds (sparrows) dataset is composed of 3 time steps with 69 birds in total. The model successfully confirms the analysis in Shizuka et al. \cite{shizuka-etal-2014} that the sparrow community is stable. The model is also applied to Indian equids (onagers) dataset by aggregating interactions of 23 onagers monthly from February 2003 to May 2003 and reveals hierarchical social integration process. \\

On the other hand, under the setting of constant community membership across different time steps and varying connectivity parameters, Bhattacharyya and Chatterjee \cite{bhattacharyya-chatterjee-2017} propose spectral clustering algorithm for dynamic stochastic blockmodel which guarantees consistency of community detection.

\subsubsection{Dynamic Mixed Membership Stochastic Blockmodel}

Xing et al. \cite{xing-etal-2010} and Ho et al. \cite{ho-etal-2011} propose dynamic extensions of the static mixed membership stochastic blockmodel using a state space model for the time-varying parameters of the priors, both for the mixed membership vector of a node and the connectivity behavior. Their methods model the role of each node as a dynamic mixed membership vector that allows nodes to behave differently over time as well as carry out different roles when interacting with different nodes.\\

Xing et al. \cite{xing-etal-2010} consider a temporal series of networks $ {Y} = \{Y_1, \ldots, Y_T\}$ where $Y_t = \{y_{ijt}\}_{i,j = 1}^{N}$ is the set of edges at time $t$ between a fixed set of $N$ nodes. In the static mixed membership stochastic blockmodels, Airoldi et al. \cite{airoldi2008mixed} employ a simple Dirichlet prior because it is conjugate to the multinomial distribution over every latent membership label $\{\vec{z}_{i \rightarrow \cdot}, \vec{z}_{i \leftarrow \cdot}\}$ defined by relevant $\vec{\pi}_{i}$. However, to model temporal dynamics of the node roles and capture the non-trivial correlations between different roles at the same time, the authors employ logistic normal distribution. 
Logistic normal distribution, $LN(\mu_t, \Sigma_t)$ is used for the prior for the mixed membership vectors, $\vec{\pi}_{it}$ and another logistic normal distribution, $LN(\eta_t, S_t)$ is used for the prior for the entries of the connectivity matrix, $\beta_{ght}$. \\

Their basic model structure is based on the state space model, which defines a linear dynamic transformation of the mixed membership priors over adjacent time points:
\begin{equation*}
\vec{\mu}_t= A\vec{\mu}_{t-1}+ \vec{\omega}_t,
\end{equation*}
where $\vec{\mu}_t$ represents the mean parameter of the prior distribution of the transformed mixed membership vectors of all vertices at time $t$, and $\vec{\omega}_t \sim N(0, \Phi)$ represents normal transition noise for the mixed membership prior, and the transition matrix $A$ shapes the trajectory of temporal transformation of the prior. Also, the model assigns similar assumption for $\eta_t$ such that
\begin{equation*}
\eta_t = b\eta_{t-1} + \xi_t,
\end{equation*}
where $b$ is scalar and $\eta_t \sim N(0, \psi)$. \\

The dynamic mixed membership stochastic blockmodel thus consists of three components: a state space model for mixed membership vector, a state space model for entries of connectivity matrix, and a logistic normal mixed membership stochastic blockmodel for the networks. The first two components explain the temporal dynamics while the third component models the generative process of the network at each time point. Following is an outline of the generative process.
\begin{itemize}
    \item State space model for mixed membership prior:
    \begin{itemize}
        \item $\vec{\mu}_1$ $\sim$ $N(\nu, \Phi)$, sample the means of the mixed membership prior at $t=1$.
        \item $\vec{\mu}_t$ $\sim$ $N(A\vec{\mu}_{t-1}, \Phi)$, sample the means of the mixed membership prior at timepoints $t=2, \ldots, T$.
    \end{itemize}
    \item State space model for connectivity matrix:
    \begin{itemize}
        \item $\eta_{gh1}$ $\sim$ $N(\iota, \psi)$, sample the means of the entries of connectivity matrix prior between $g$ and $h$ at $t=1$.
        \item $\eta_{ght}$ $\sim$ $N(b\eta_{gh(t-1)}, \psi)$, sample the means of the entries of connectivity matrix prior between $g$ and $h$ at $t=2, \ldots, T$.
        \item $\beta_{ght}$ $\sim$ $LN(\eta_t, S_t)$, sample the entries of connectivity matrix between $g$ and $h$ at $t=1, 2, \ldots, T$.
    \end{itemize}
    \item Logistic normal mixed membership stochastic blockmodel:
    \begin{itemize}
        \item $\vec{\pi}_{it}$ $\sim$ $LN(\vec{\mu}_t, \Sigma_t)$, sample a $C$-dimensional mixed membership vector for each node $i=1, \ldots, N$, at $t=1,2, \ldots, T$.
        \item $\vec{z}_{i \rightarrow j,t}$ $\sim$ $\mbox{Multinomial}(\vec{\pi}_{it})$, sample membership indicator for the initiator for each node $i=1, \ldots, N$, at $t=1,2, \ldots, T$.
        \item $\vec{z}_{i \leftarrow j,t}$ $\sim$ $\mbox{Multinomial}(\vec{\pi}_{jt})$, sample membership indicator for the receiver for each node $i=1, \ldots, N$, at $t=1,2, \ldots, T$.
        \item $e_{ijt}$  $\sim$ $\mbox{Bernoulli}(\vec{z}_{i \rightarrow j,t'}B_t\vec{z}_{i \leftarrow j,t})$, sample the edges between nodes $i$ and $j$ for $1 < i \neq j < N$, at $t=1,2, \ldots, T$.
    \end{itemize}
\end{itemize}

The joint probability of the data ${Y}$ and the latent variables $\{\vec{\pi}_{1:N,t},Z_{\rightarrow,t},Z_{\leftarrow,t}\}$ can be written in similar form as the static mixed membership stochastic blockmodel. \\

For the posterior inference, Laplace variational approximation scheme based on the generalized means field (GMF) approximation is used to infer the latent variables and estimate the model parameters.\\

This model is applied to three datasets. The Sampson's monk dataset contains the liking relationship among 18 monks over three time points. The result of the model is consistent with previous works except for one controversial person, Mark. The model is also applied to a subset of Enron email dataset of 151 persons from 2001 and discovers five distinct roles of actors. They visualize and track the trajectory of the mixed membership vector for an individual to understand how each evolves in his or her role. The third example is studying the evolving gene network of fruit flies. The Drosophila melanogaster gene network dataset contains 22 networks at different time points across various developmental stages. The model discovers that many genes exhibit a sharp transition regarding their roles near the end of the embryonic stage. They further examine 45 ontological groups and find an overall pattern that each role consists of genes with a variety of functions, and the functional composition of each role varies across time. \\

Different from Xing et al. \cite{xing-etal-2010} which employs a time-evolving logistic normal distribution on all networks nodes, Ho et al. \cite{ho-etal-2011} generalize the prior on nodes to be a mixture of time-evolving logistic normal distributions. This mixture prior is multi-modal and captures correlations between roles, allowing it to fit complex data densities that the unimodal prior cannot. \\

The state space model, which defines a linear dynamic transformation of the mixed membership priors, now contains $N$ distinct trajectories over adjacent time points,
\begin{equation*}
\vec{\mu}_{ct} = A\vec{\mu}_{c(t-1)} + \vec{\omega}_{ct}, \quad c=1, 2, \ldots, C
\end{equation*}
where $A$ is a transition matrix and $\omega_{ct}$ $\sim$ $N(0, \Phi)$ represents the normal transition noise. Now each mixed membership vector $\vec{\pi}_{it}$ is drawn from one of the $C$ trajectories $\vec{\mu}_{ct}$. The choice of trajectory for $\vec{\pi}_{it}$ is given by the indicator $z_{it}$, which is drawn from some prior. For simplicity, they use a single multinomial prior with the parameter $\delta$ for all $z_{it}$. Notice that $z_{it}$ can change over time, allowing nodes to switch clusters if that would fit the data better. Given $z_{it}$, mixed membership vector $\vec{\pi}_{it}$ is sampled from $LN(\vec{\mu}_{z_{it},t}, \Sigma_{z_{it},t})$. Once $\{\vec{\pi}_{it}\}_{i=1}^{N}$ have been drawn for some timepoint $t$, the remaining part follows the generative process of Xing et al. \cite{xing-etal-2010}. They also propose a similar variational EM algorithm for parameter learning and latent variable inference. \\ 

The authors analyze the United States 109th Congress voting records by generating time-varying networks containing 100 senators and eight time points corresponding to 3-month periods from Jan 1st, 2005 to Dec 31st, 2006. The analysis not only finds the two clusters capturing party affiliations but also identifies outliers, for example, a senator who shifted from the Democrat cluster to the Republican cluster.

\section{Conclusions and Discussions} \label{sec:discussion}
 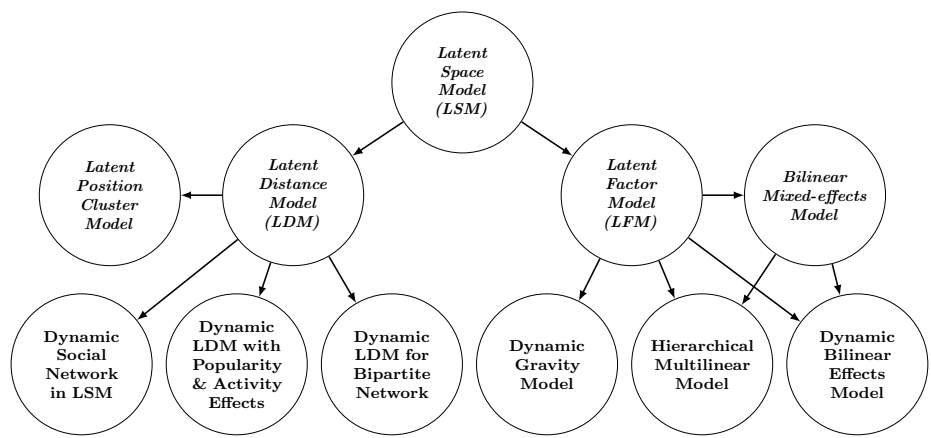
\begin{figure}[!b]
 	\centering
 	\scalebox{0.75}{
 		\begin{tikzpicture}[
 		block/.style = {draw, fill=white!30, align=center, anchor=west,
 			minimum height=0.5cm, inner sep=0},
 		ball/.style = {circle, draw, align=center, anchor=north, minimum size = 2.5cm, inner sep=-3}]
 		\tikzstyle{connect}=[-latex, thick]
 \node[ball,text width=2.1cm] (PAEM) at (1,-5) {\bf Dynamic LDM with Popularity \& Activity Effects};
 	\node[ball,text width=2.1cm] (BNM) at (3.75,-5) {\bf Dynamic LDM for Bipartite Network};
 		\node[ball,text width=2.1cm] (LSM) at (5,0) {\bf  \textit{Latent\\Space\\Model
 			\\(LSM)}};
 		\node[ball,text width=2.1cm] (LPCM) at (-1.25,-2) {\bf \textit{Latent\\Position\\Cluster\\Model}};
 		\node[ball,text width=2.1cm] (BMEM) at (11.25,-2) {\bf \textit{Bilinear\\Mixed-effects\\Model}};
 		\node[ball,text width=2.1cm] (LDM) at (2,-2) {\bf \textit{Latent\\Distance\\Model\\(LDM)}};
 		\node[ball,text width=2.1cm] (LFM) at (8,-2) {\bf \textit{Latent\\Factor\\Model\\(LFM)}};
 		\node[ball,text width=2.1cm] (DSNL) at (-1.75,-5) {\bf Dynamic \\Social Network \\in LSM};
 		\node[ball,text width=2.1cm] (DGM) at (6.5,-5) {\bf Dynamic Gravity \\Model};
 		\node[ball,text width=2.1cm] (HMM) at (9.25,-5) {\bf Hierarchical Multilinear Model};
 		\node[ball,text width=2.1cm] (DBEM) at (12,-5) {\bf Dynamic Bilinear\\Effects \\Model};
 		
 		\path (LSM) edge [connect] (LDM)
 		(LDM) edge [connect] (LPCM)
 		(LSM) edge [connect] (LFM)
 		(LFM) edge [connect] (BMEM)
 		(LDM) edge [connect] (DSNL)
 		(LDM) edge [connect] (PAEM)
 		(LDM) edge [connect] (BNM)
 		(LFM) edge [connect] (DGM)
 		(LFM) edge [connect] (HMM)
 		(BMEM) edge [connect] (HMM)
 		(BMEM) edge [connect] (DBEM)
 		(LFM) edge [connect] (DBEM);
 		\end{tikzpicture}}
 	\caption{Network summarizing the relations between models discussed in our review of latent space models (LSM): static latent space models (\textit{red}) and dynamic latent space models (\textit{blue}). Arrows indicate inspiration or influence of the model at the source on the model at
 		the target.}
 	\label{fig:LSM}
 \end{figure}

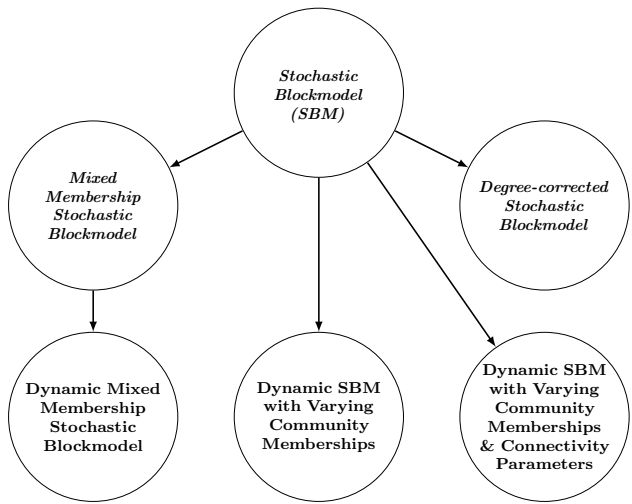
\begin{figure}[!t]
	\centering
	\scalebox{0.75}{
		\begin{tikzpicture}[
		block/.style = {draw, fill=white!30, align=center, anchor=west,
			minimum height=0.65cm, inner sep=0},
		ball/.style = {circle, draw, align=center, anchor=north, minimum size = 3cm,inner sep=-3}]
		\tikzstyle{connect}=[-latex, thick]
		\node[ball,text width=2.5cm] (SBM) at (5,0) {\bf \\ \textit{Stochastic\\Blockmodel\\(SBM)}};
		\node[ball,text width=2.5cm] (MMBM) at (1,-2) {\bf \textit{Mixed\\ Membership\\Stochastic\\Blockmodel}};
		\node[ball,text width=2.5cm] (DCM) at (9,-2) {\bf \textit{Degree-corrected\\Stochastic\\Blockmodel}};
		\node[ball,text width=2.5cm] (VCMM) at (5,-5.75) {\bf Dynamic SBM with Varying Community Memberships};
		\node[ball,text width=2.5cm] (CPM) at (9,-5.75) {\bf Dynamic SBM with Varying Community Memberships \& Connectivity Parameters};
		\node[ball,text width=2.5cm] (DMMM) at (1,-5.75) {\bf Dynamic Mixed Membership Stochastic Blockmodel};
		\path (SBM) edge [connect] (MMBM)
		 (SBM) edge [connect] (DCM)
		  (SBM) edge [connect] (VCMM)
		   (SBM) edge [connect] (CPM)
		        (MMBM) edge [connect] (DMMM);
		\end{tikzpicture}}
	\caption{Network summarizing the relations between models discussed in our review of stochastic blockmodels (SBM): static stochastic blockmodels (\textit{red}) and dynamic stochastic models (\textit{blue}). Arrows indicate inspiration or influence of the model at the source on the model at the target.}
	\label{fig:SBM}
\end{figure}

The literature on statistical network modeling may be divided into various lines of research, such as the development of network statistics for ERGMs, inference methods for SAOMs, and asymptotic network properties in physics-based models. Among various possibilities, in this paper, we mainly focus on the review of dynamic network models with latent variables---models that assume and estimate latent structures of networks---in detail. Latent space models assume the existence of unobserved network embedding into a low-dimensional space, and stochastic blockmodels assume that the nodes within same blocks are structurally equivalence. From these, we make a distinction between the latent space models (LSM) and stochastic blockmodels (SBM) and outline the evolution of each class of models from static to dynamic models, as well as their explicit connections. As a review of network models, to conclude, we provide two mini directed networks to summarize the relationships of the models discussed in previous sections, with latent space models in Figure \ref{fig:LSM} and stochastic blockmodels in Figure \ref{fig:SBM}. In both figures, static models are in italic letters, with the names of nodes corresponding to the title of each section. A directed link represents the sender model motivates the development of the receiver model, or the receiver is a dynamic extension of the sender.\newline

 Despite recent advances in statistical and computational methods of dynamic network modeling, there remain several challenges across the models described in previous sections. Some (e.g., computation, visualization) are challenges for network modeling in general, while others (e.g., lack of continuous-time models, identifiability) are only applicable to dynamic latent space and/or stochastic blockmodels. In this section, we briefly sketch out the three major issues in modeling dynamic networks with latent variables.\\

Computation, the most common type of problem for network modeling in general, is particularly an issue for latent variable models due to the estimation of the additional parameters. For example, the evaluation of likelihood equations for static and dynamic latent space models require the calculation of a distance matrix given the complete set of latent positions at a computational and storage cost of $O(N^2)$ \cite{rastelli2016properties} and $O(T\times N^2)$, respectively. This makes the estimation of latent space models impractical for networks larger than a few thousand nodes, and as a result, most of their applications have been limited to relatively small networks, typically less than 500 nodes. As the existing software packages for other well-known frameworks including temporal ERGMs \cite{hanneke2010discrete,krivitsky2014separable,lee-etal-2017,agarwal-etal-2017} or SAOMs \cite{snijders1996stochastic,snijders2010introduction} can efficiently fit networks of size up to a few thousand nodes, there is a strong need for the development of efficient statistical software to fit these models. Recently, new estimation methods \cite{ma2017exploration,raftery2012fast,rastelli2016properties,rastelli2018computationally} are considered for latent space models, so we expect a bright prospect to achieve a substantial reduction in computing time. In the case of stochastic blockmodels, the computation is especially non-trivial for large networks, since the optimization over all possible label assignments is NP-hard (i.e., non-deterministic polynomial-time hardness). With the help of variational methods \cite{blei2017variational}, stochastic blockmodels can be fit for
thousands of nodes. It is still a challenge to fit blockmodels to large networks with millions or tens of millions of nodes. \\

In general, latent variable models allow for better understanding of networks where it is believed to contain hidden structure. As such, visualization is one of the biggest advantages of those models due to their model-based nature such as Euclidean distances between nodes and clustering of the nodes. At the same time, there also remain limitations and open problems in visualizing the latent variables. For example, latent space models are usually visualized in 2D or 3D plots, but not in higher dimensions. Furthermore, when a dynamic network spans over a large number of time points (e.g., $T \geq 50$) or based on continuous-time scale, it is also difficult to come up with plots that can effectively show the movement of latent positions.\\

Finally, motivated by the rise of large-scale online social networks, a variety of continuous-time network models have been introduced, mostly built upon survival and event history analysis \cite{hunter2011dynamic,perry2013point,vu2011continuous}. In contrast, most of the dynamic latent variable models\footnote{One exception is the dynamic bilinear effects model in Section \ref{subsubsec:Dunson}, however, the illustrated example is still the discrete (quarterly) network data with $T = 39$.} introduced in this paper focus on learning discrete-time networks or the discrete realization of a continuous process \cite{durante2014}, where multiple cross-sectional ``snapshots" of the network are recorded at discrete time points. Considering the growing number of network data recorded on a continuous-time scale, developing the latent variable models that take advantage of these temporal
granularity might be a promising future research direction.

\bibliographystyle{apalike}

\begin{thebibliography}{}

\end{thebibliography}


\begin{thebibliography}{}

\bibitem[Agarwal et~al., 2018]{agarwal-etal-2017}
Agarwal, A., Lee, K., and Xue, L. (2018).
\newblock Temporal exponential-family random graph models with time-evolving
  latent block structure for dynamic networks.
\newblock {\em Technical report, Penn State Univeristy}.

\bibitem[Airoldi et~al., 2008]{airoldi2008mixed}
Airoldi, E.~M., Blei, D.~M., Fienberg, S.~E., and Xing, E.~P. (2008).
\newblock Mixed membership stochastic blockmodels.
\newblock {\em Journal of Machine Learning Research}, 9(Sep):1981--2014.

\bibitem[Amini et~al., 2013]{amini-etal-2013}
Amini, A.~A., Chen, A., Bickel, P.~J., and Levina, E. (2013).
\newblock Pseudo-likelihood methods for community detection in large sparse
  networks.
\newblock {\em The Annals of Statistics}, 31(4):2097--2122.

\bibitem[Amini and Levina, 2018]{amini-levina-2018}
Amini, A.~A. and Levina, E. (2018).
\newblock On semidefinite relaxations for the block model.
\newblock {\em The Annals of Statistics}, 46(1):149--179.

\bibitem[Arbeitman et~al., 2002]{arbeitman-etal-2002}
Arbeitman, M.~N., Furlong, E.~E., Imam, F., Johnson, E., Null, B.~H., Baker,
  B.~S., Krasnow, M.~A., Scott, M.~P., Davis, R.~W., and White., K.~P. (2002).
\newblock Gene expression during the life cycle of drosophila melanogaster.
\newblock {\em Science}, 297(5590):2270--2275.

\bibitem[Asur and Parthasarathy, 2007]{asur-parthasarathy-2007}
Asur, S. and Parthasarathy, S. (2007).
\newblock An event-based framework for characterizing the evolutionary behavior
  of interaction graphs.
\newblock {\em Proceedings of the 13th ACM SIGKDD International Conference on
  Knowledge Discovery and Data Mining}, pages 913--921.

\bibitem[Banfield and Raftery, 1993]{mbc93}
Banfield, J. and Raftery, A. (1993).
\newblock Model-based gaussian and non-gaussian clustering.
\newblock {\em Biometrics}, 49:803--821.

\bibitem[Bhattacharyya and Chatterjee, 2017]{bhattacharyya-chatterjee-2017}
Bhattacharyya, S. and Chatterjee, S. (2017).
\newblock Spectral clustering for dynamic stochastic block model.
\newblock {\em Working Paper}.

\bibitem[Blei et~al., 2017]{blei2017variational}
Blei, D.~M., Kucukelbir, A., and McAuliffe, J.~D. (2017).
\newblock Variational inference: A review for statisticians.
\newblock {\em Journal of the American Statistical Association}, to appear.

\bibitem[Chang and Blei, 2009]{chang2009relational}
Chang, J. and Blei, D.~M. (2009).
\newblock Relational topic models for document networks.
\newblock In {\em International Conference on Artificial Intelligence and
  Statistics}, pages 81--88.

\bibitem[Chi et~al., 2007]{chi-etal-2007}
Chi, Y., Song, X., Zhou, D., Hino, K., and Tseng, B.~L. (2007).
\newblock Evolutionary spectral clustering by incorporating temporal
  smoothness.
\newblock {\em In Proceedings of the 13th ACM SIGKDD International Conference
  on Knowledge Discovery and Data Mining}, pages 153--162.

\bibitem[Desmarais and Cranmer, 2017]{desmarais2017statistical}
Desmarais, B.~A. and Cranmer, S.~J. (2017).
\newblock Statistical inference in political networks research.
\newblock {\em arXiv preprint arXiv:1703.02870}.

\bibitem[Durante and Dunson, 2014a]{durante2014bayesian}
Durante, D. and Dunson, D.~B. (2014a).
\newblock Bayesian dynamic financial networks with time-varying predictors.
\newblock {\em Statistics \& Probability Letters}, 93:19--26.

\bibitem[Durante and Dunson, 2014b]{durante2014bayesian2}
Durante, D. and Dunson, D.~B. (2014b).
\newblock Bayesian logistic gaussian process models for dynamic networks.
\newblock In {\em Artificial Intelligence and Statistics}, pages 194--201.

\bibitem[Durante and Dunson, 2014c]{durante2014}
Durante, D. and Dunson, D.~B. (2014c).
\newblock Nonparametric bayes dynamic modeling of relational data.
\newblock {\em Biometrika}, 101(4):883--898.

\bibitem[Eagle et~al., 2009]{eagle-etal-2009}
Eagle, N., Pentland, A.~S., and Lazer, D. (2009).
\newblock Inferring friendship network structure by using mobile phone data.
\newblock {\em Proceedings of the National Academy of Sciences},
  106(36):15274--15278.

\bibitem[Fletcher et~al., 2011]{fletcher2011social}
Fletcher, R.~J., Acevedo, M.~A., Reichert, B.~E., Pias, K.~E., and Kitchens,
  W.~M. (2011).
\newblock Social network models predict movement and connectivity in ecological
  landscapes.
\newblock {\em Proceedings of the National Academy of Sciences},
  108(48):19282--19287.

\bibitem[Freeman, 2003]{freeman-2003}
Freeman, L.~C. (2003).
\newblock {\em Finding social groups: A meta-analysis of the southern women
  data}.
\newblock New York: National Academies Press.

\bibitem[Friel et~al., 2016]{friel2016interlocking}
Friel, N., Rastelli, R., Wyse, J., and Raftery, A.~E. (2016).
\newblock Interlocking directorates in irish companies using a latent space
  model for bipartite networks.
\newblock {\em Proceedings of the National Academy of Sciences},
  113(24):6629--6634.

\bibitem[Girvan and Newman, 2002]{girvan-newman-2002}
Girvan, M. and Newman, M.~E. (2002).
\newblock Community structure in social and biological networks.
\newblock {\em Proceedings of the National Academy of Sciences},
  99(12):7821--7826.

\bibitem[Goldenberg et~al., 2010]{goldenberg2010survey}
Goldenberg, A., Zheng, A.~X., Fienberg, S.~E., Airoldi, E.~M., et~al. (2010).
\newblock A survey of statistical network models.
\newblock {\em Foundations and Trends{\textregistered} in Machine Learning},
  2(2):129--233.

\bibitem[Gu{\'e}don and Vershynin, 2016]{guedon-vershynin-2016}
Gu{\'e}don, O. and Vershynin, R. (2016).
\newblock Community detection in sparse networks via grothendieck's inequality.
\newblock {\em Probability Theory and Related Fields}, 165(3-4):1025--1049.

\bibitem[Handcock et~al., 2007]{handcock2007model}
Handcock, M.~S., Raftery, A.~E., and Tantrum, J.~M. (2007).
\newblock Model-based clustering for social networks.
\newblock {\em Journal of the Royal Statistical Society: Series A (Statistics
  in Society)}, 170(2):301--354.

\bibitem[Hanneke et~al., 2010]{hanneke2010discrete}
Hanneke, S., Fu, W., Xing, E.~P., et~al. (2010).
\newblock Discrete temporal models of social networks.
\newblock {\em Electronic Journal of Statistics}, 4:585--605.

\bibitem[Henry et~al., 2016]{henry2016modeling}
Henry, T., Banks, D., Chai, C., and Owens-Oas, D. (2016).
\newblock Modeling community structure and topics in dynamic text networks.
\newblock {\em arXiv preprint arXiv:1610.05756}.

\bibitem[Ho et~al., 2011]{ho-etal-2011}
Ho, Q., Song, L., and Xing, E. (2011).
\newblock Evolving cluster mixed-membership blockmodel for time-varying
  networks.
\newblock {\em Journal of Machine Learning Research: Workshop and Conference
  Proceedings}, 15:342--350.

\bibitem[Hoff et~al., 2013]{hoff2013likelihoods}
Hoff, P., Fosdick, B., Volfovsky, A., and Stovel, K. (2013).
\newblock Likelihoods for fixed rank nomination networks.
\newblock {\em Network Science}, 1(3):253--277.

\bibitem[Hoff, 2005]{hoff2005bilinear}
Hoff, P.~D. (2005).
\newblock Bilinear mixed-effects models for dyadic data.
\newblock {\em Journal of the American Statistical Association},
  100(469):286--295.

\bibitem[Hoff, 2008]{hoff2008modeling}
Hoff, P.~D. (2008).
\newblock Modeling homophily and stochastic equivalence in symmetric relational
  data.
\newblock In {\em Advances in Neural Information Processing Systems}, pages
  657--664.

\bibitem[Hoff, 2009]{hoff2009multiplicative}
Hoff, P.~D. (2009).
\newblock Multiplicative latent factor models for description and prediction of
  social networks.
\newblock {\em Computational and Mathematical Organization Theory},
  15(4):261--272.

\bibitem[Hoff, 2011]{hoff2011hierarchical}
Hoff, P.~D. (2011).
\newblock Hierarchical multilinear models for multiway data.
\newblock {\em Computational Statistics \& Data Analysis}, 55(1):530--543.

\bibitem[Hoff, 2015a]{hoff2015dyadic}
Hoff, P.~D. (2015a).
\newblock Dyadic data analysis with amen.
\newblock {\em arXiv preprint arXiv:1506.08237}.

\bibitem[Hoff, 2015b]{hoff2015multilinear}
Hoff, P.~D. (2015b).
\newblock Multilinear tensor regression for longitudinal relational data.
\newblock {\em The Annals of Applied Statistics}, 9(3):1169.

\bibitem[Hoff et~al., 2011]{hoff2011separable}
Hoff, P.~D. et~al. (2011).
\newblock Separable covariance arrays via the tucker product, with applications
  to multivariate relational data.
\newblock {\em Bayesian Analysis}, 6(2):179--196.

\bibitem[Hoff et~al., 2002]{hoff2002latent}
Hoff, P.~D., Raftery, A.~E., and Handcock, M.~S. (2002).
\newblock Latent space approaches to social network analysis.
\newblock {\em Journal of the American Statistical Association},
  97(460):1090--1098.

\bibitem[Hoff and Ward, 2004]{hoff2004modeling}
Hoff, P.~D. and Ward, M.~D. (2004).
\newblock Modeling dependencies in international relations networks.
\newblock {\em Political Analysis}, 12(2):160--175.

\bibitem[Holland et~al., 1983]{holland-etal-1983}
Holland, P.~W., Laskey, K.~B., and Leinhardt, S. (1983).
\newblock Stochastic blockmodels: First steps.
\newblock {\em Social Networks}, 5(2):109--137.

\bibitem[Hunter et~al., 2011]{hunter2011dynamic}
Hunter, D., Smyth, P., Vu, D.~Q., and Asuncion, A.~U. (2011).
\newblock Dynamic egocentric models for citation networks.
\newblock In {\em Proceedings of the 28th International Conference on Machine
  Learning (ICML-11)}, pages 857--864.

\bibitem[Ji and Jin., 2016]{ji-jin-2014}
Ji, P. and Jin., J. (2016).
\newblock Coauthorship and citation networks for statisticians.
\newblock {\em The Annals of Applied Statistics}, 10(4):1779--1812.

\bibitem[Jin et~al., 2017]{jin-etal-2017}
Jin, J., Ke, Z.~T., and Luo, S. (2017).
\newblock Estimating network memberships by simplex vertex hunting.
\newblock {\em arXiv preprint arXiv:1708.07852}.

\bibitem[Karrer and Newman, 2011]{karrer-newman-2011}
Karrer, B. and Newman, M. E.~J. (2011).
\newblock Stochastic blockmodels and community structure in networks.
\newblock {\em Physical Review E}, 83(1):016107.

\bibitem[Kolaczyk, 2009]{kolaczyk2009statistical}
Kolaczyk, E.~D. (2009).
\newblock {\em Statistical Analysis of Network Data: Methods and Models}.
\newblock Springer Science \& Business Media.

\bibitem[Krafft et~al., 2012]{krafft2012topic}
Krafft, P., Moore, J., Desmarais, B., and Wallach, H.~M. (2012).
\newblock Topic-partitioned multinetwork embeddings.
\newblock In {\em Advances in Neural Information Processing Systems}, pages
  2807--2815.

\bibitem[Krivitsky and Handcock, 2008]{latentnet2}
Krivitsky, P.~N. and Handcock, M.~S. (2008).
\newblock Fitting position latent cluster models for social networks with
  latentnet.
\newblock {\em Journal of Statistical Software}, 24(5).

\bibitem[Krivitsky and Handcock, 2014]{krivitsky2014separable}
Krivitsky, P.~N. and Handcock, M.~S. (2014).
\newblock A separable model for dynamic networks.
\newblock {\em Journal of the Royal Statistical Society: Series B (Statistical
  Methodology)}, 76(1):29--46.

\bibitem[Krivitsky and Handcock, 2015]{latentnet}
Krivitsky, P.~N. and Handcock, M.~S. (2015).
\newblock {\em latentnet: Latent Position and Cluster Models for Statistical
  Networks}.
\newblock The Statnet Project (\url{http://www.statnet.org}).
\newblock R package version 2.7.1.

\bibitem[Krivitsky et~al., 2009]{krivitsky2009representing}
Krivitsky, P.~N., Handcock, M.~S., Raftery, A.~E., and Hoff, P.~D. (2009).
\newblock Representing degree distributions, clustering, and homophily in
  social networks with latent cluster random effects models.
\newblock {\em Social Networks}, 31(3):204--213.

\bibitem[Lee et~al., 2017]{lee-etal-2017}
Lee, K., Xue, L., and Hunter, D.~R. (2017).
\newblock Model-based clustering of time-evolving networks through temporal
  exponential-family random graph models.
\newblock {\em Technical report, Penn State Univeristy}.

\bibitem[Leibenstein, 1966]{leibenstein1966shaping}
Leibenstein, H. (1966).
\newblock Shaping the World Economy: Suggestions for an International Economic
  Policy.

\bibitem[Lin et~al., 2008]{lin-etal-2008}
Lin, Y., Chi, Y., Zhu, S., Sundaram, H., and Tseng, B.~L. (2008).
\newblock Facetnet: a framework for analyzing communities and their evolutions
  in dynamic networks.
\newblock {\em In Proceedings of the 17th International Conference on World
  Wide Web}, pages 685--694.

\bibitem[Ma and Ma, 2017]{ma2017exploration}
Ma, Z. and Ma, Z. (2017).
\newblock Exploration of large networks via fast and universal latent space
  model fitting.
\newblock {\em arXiv preprint arXiv:1705.02372}.

\bibitem[Mao et~al., 2017]{mao-etal-2017}
Mao, X., Sarkar, P., and Chakrabarti, D. (2017).
\newblock Estimating mixed memberships with sharp eigenvector deviations.
\newblock {\em arXiv preprint arXiv:1709.00407}.

\bibitem[Matias and Miele, 2017]{matias-miele-2017}
Matias, C. and Miele, V. (2017).
\newblock Statistical clustering of temporal networks through a dynamic
  stochastic block model.
\newblock {\em Journal of the Royal Statistical Society: Series B (Statistical
  Methodology)}, 79(4):1119--1141.

\bibitem[Matias and Robin, 2014]{matias2014modeling}
Matias, C. and Robin, S. (2014).
\newblock Modeling heterogeneity in random graphs through latent space models:
  a selective review.
\newblock {\em ESAIM: Proceedings and Surveys}, 47:55--74.

\bibitem[Minhas et~al., 2016a]{minhas2016inferential}
Minhas, S., Hoff, P.~D., and Ward, M.~D. (2016a).
\newblock Inferential approaches for network analyses: Amen for latent factor
  models.
\newblock {\em arXiv preprint arXiv:1611.00460}.

\bibitem[Minhas et~al., 2016b]{minhas2016new}
Minhas, S., Hoff, P.~D., and Ward, M.~D. (2016b).
\newblock A new approach to analyzing coevolving longitudinal networks in
  international relations.
\newblock {\em Journal of Peace Research}, 53(3):491--505.

\bibitem[Newman et~al., 2011]{newman2011structure}
Newman, M., Barabasi, A.-L., and Watts, D.~J. (2011).
\newblock {\em The Structure and Dynamics of Networks}.
\newblock Princeton University Press.

\bibitem[Newman, 2004]{newman-2004}
Newman, M.~E. (2004).
\newblock Coauthorship networks and patterns of scientific collaboration.
\newblock {\em Proceedings of the National Academy of Sciences}, 101 (suppl
  1):5200--5205.

\bibitem[Nowicki and Snijders, 2001]{nowicki-2001}
Nowicki, K. and Snijders, T.~A. (2001).
\newblock Estimation and prediction for stochastic blockstructures.
\newblock {\em Journal of the American Statistical Association},
  96(455):1077--1087.

\bibitem[Perry and Wolfe, 2013]{perry2013point}
Perry, P.~O. and Wolfe, P.~J. (2013).
\newblock Point process modelling for directed interaction networks.
\newblock {\em Journal of the Royal Statistical Society: Series B (Statistical
  Methodology)}, 75(5):821--849.

\bibitem[Qin and Rohe, 2013]{qin-rohe-2013}
Qin, T. and Rohe, K. (2013).
\newblock Regularized spectral clustering under the degree-corrected stochastic
  blockmodel.
\newblock {\em Advances in Neural Information Processing Systems}, pages
  3120--3128.

\bibitem[Raftery et~al., 2012]{raftery2012fast}
Raftery, A.~E., Niu, X., Hoff, P.~D., and Yeung, K.~Y. (2012).
\newblock Fast inference for the latent space network model using a
  case-control approximate likelihood.
\newblock {\em Journal of Computational and Graphical Statistics},
  21(4):901--919.

\bibitem[Rastelli et~al., 2016]{rastelli2016properties}
Rastelli, R., Friel, N., and Raftery, A.~E. (2016).
\newblock Properties of latent variable network models.
\newblock {\em Network Science}, 4(4):407--432.

\bibitem[Rastelli et~al., 2018]{rastelli2018computationally}
Rastelli, R., Maire, F., and Friel, N. (2018).
\newblock Computationally efficient inference for latent position network
  models.
\newblock {\em arXiv preprint arXiv:1804.02274}.

\bibitem[Rohe et~al., 2011]{rohe-etal-2011}
Rohe, K., Chatterjee, S., and Yu, B. (2011).
\newblock Spectral clustering and the high-dimensional stochastic blockmodel.
\newblock {\em The Annals of Statistics}, pages 1878--1915.

\bibitem[Rubenstein et~al., 2015]{rubenstein-etal-2015}
Rubenstein, D.~I., Sundaresan, S.~R., Fischhoff, I.~R., Tantipathananandh, C.,
  and Berger-Wolf, T.~Y. (2015).
\newblock Similar but different: Dynamic social network analysis highlights
  fundamental differences between the fission-fusion societies of two equid
  species, the onager and grevy's zebra.
\newblock {\em PloS ONE}, 10(10).

\bibitem[Sarkar and Moore, 2005]{sarkar2005dynamic}
Sarkar, P. and Moore, A.~W. (2005).
\newblock Dynamic social network analysis using latent space models.
\newblock {\em ACM SIGKDD Explorations Newsletter}, 7(2):31--40.

\bibitem[Sarkar et~al., 2007]{sarkar2007latent}
Sarkar, P., Siddiqi, S.~M., and Gordon, G.~J. (2007).
\newblock A latent space approach to dynamic embedding of co-occurrence data.
\newblock In {\em AISTATS}, pages 420--427.

\bibitem[Sewell and Chen, 2015]{sewell2015latent}
Sewell, D.~K. and Chen, Y. (2015).
\newblock Latent space models for dynamic networks.
\newblock {\em Journal of the American Statistical Association},
  110(512):1646--1657.

\bibitem[Sewell and Chen, 2016]{sewell2016latent}
Sewell, D.~K. and Chen, Y. (2016).
\newblock Latent space models for dynamic networks with weighted edges.
\newblock {\em Social Networks}, 44:105--116.

\bibitem[Shizuka et~al., 2014]{shizuka-etal-2014}
Shizuka, D., Chaine, A.~S., Anderson, J., Johnson, O., Laursen, I.~M., and
  Lyon, B.~E. (2014).
\newblock Across-year social stability shapes network structure in wintering
  migrant sparrows.
\newblock {\em Ecology Letters}, 17(8):998--1007.

\bibitem[Snijders, 1996]{snijders1996stochastic}
Snijders, T.~A. (1996).
\newblock Stochastic actor-oriented models for network change.
\newblock {\em Journal of Mathematical Sociology}, 21(1-2):149--172.

\bibitem[Snijders, 2011]{snijders2011statistical}
Snijders, T.~A. (2011).
\newblock Statistical models for social networks.
\newblock {\em Annual Review of Sociology}, 37.

\bibitem[Snijders and Nowicki, 1997]{snijders-1997}
Snijders, T.~A. and Nowicki, K. (1997).
\newblock Estimation and prediction for stochastic blockmodels for graphs with
  latent block structure.
\newblock {\em Journal of Classification}, 14:75--100.

\bibitem[Snijders et~al., 2010]{snijders2010introduction}
Snijders, T.~A., Van~de Bunt, G.~G., and Steglich, C.~E. (2010).
\newblock Introduction to stochastic actor-based models for network dynamics.
\newblock {\em Social Networks}, 32(1):44--60.

\bibitem[Sweet and Junker, 2011]{sweet2011modeling}
Sweet, T.~M. and Junker, B. (2011).
\newblock Modeling intervention effects on social networks in education
  research.
\newblock {\em Educational Evaluation and Policy Analysis}, 30:203--235.

\bibitem[Sweet et~al., 2013]{sweet2013hierarchical}
Sweet, T.~M., Thomas, A.~C., and Junker, B.~W. (2013).
\newblock Hierarchical network models for education research: Hierarchical
  latent space models.
\newblock {\em Journal of Educational and Behavioral Statistics},
  38(3):295--318.

\bibitem[Vivar and Banks, 2012]{vivar2012models}
Vivar, J.~C. and Banks, D. (2012).
\newblock Models for networks: a cross-disciplinary science.
\newblock {\em Wiley Interdisciplinary Reviews: Computational Statistics},
  4(1):13--27.

\bibitem[Vu et~al., 2011]{vu2011continuous}
Vu, D.~Q., Hunter, D., Smyth, P., and Asuncion, A.~U. (2011).
\newblock Continuous-time regression models for longitudinal networks.
\newblock In {\em Advances in Neural Information Processing Systems}, pages
  2492--2500.

\bibitem[Ward et~al., 2013]{ward2013gravity}
Ward, M.~D., Ahlquist, J.~S., and Rozenas, A. (2013).
\newblock Gravity's rainbow: A dynamic latent space model for the world trade
  network.
\newblock {\em Network Science}, 1(1):95--118.

\bibitem[Ward and Hoff, 2007]{ward2007persistent}
Ward, M.~D. and Hoff, P.~D. (2007).
\newblock Persistent patterns of international commerce.
\newblock {\em Journal of Peace Research}, 44(2):157--175.

\bibitem[Ward et~al., 2007]{ward2007disputes}
Ward, M.~D., Siverson, R.~M., and Cao, X. (2007).
\newblock Disputes, democracies, and dependencies: A reexamination of the
  kantian peace.
\newblock {\em American Journal of Political Science}, 51(3):583--601.

\bibitem[Ward et~al., 2011]{ward2011network}
Ward, M.~D., Stovel, K., and Sacks, A. (2011).
\newblock Network analysis and political science.
\newblock {\em Annual Review of Political Science}, 14:245--264.

\bibitem[Xing et~al., 2010]{xing-etal-2010}
Xing, E.~P., Fu, W., and Song, L. (2010).
\newblock A state-space mixed membership blockmodel for dynamic network
  tomography.
\newblock {\em The Annals of Applied Statistics}, 4(2).

\bibitem[Xu and Hero, sing]{xu-hero-2014}
Xu, K.~S. and Hero, A.~O. (2014).
\newblock Dynamic stochastic blockmodels for time-evolving social networks.
\newblock {\em IEEE Journal of Selected Topics in Signal
  Processing}, 8(4):552--562.

\bibitem[Yang et~al., 2011]{yang-etal-2011}
Yang, T., Chi, Y., Zhu, S., Gong, Y., and Jin, R. (2011).
\newblock Detecting communities and their evolutions in dynamic social
  networks-a bayesian approach.
\newblock {\em Machine learning}, 82(2):157--189.

\bibitem[Zhao et~al., 2012]{zhao-etal-2012}
Zhao, Y., Levina, E., and Zhu, J. (2012).
\newblock Consistency of community detection in networks under degree-corrected
  stochastic block models.
\newblock {\em The Annals of Statistics}, 40(4):2266--2292.

\end{thebibliography}

\section*{Appendix: Dynamic Network Data Sources}		
We provide a list of data sets mentioned in this paper and the current version of data sources. In case of multiple sources, we refer to multiple websites or a specific section of the paper in which the data sources are illustrated in detail.
\begin{center}
	\scalebox{0.625}{	
		\begin{tabular}{ |c|c| c|c|}
			\hline
			Data & Article & Section & Source\\
			\hline\hline
			NIPS data &\cite{sarkar2005dynamic,sarkar2007latent}& 2.2.1&\footnotesize\url{http://ai.stanford.edu/~gal/data.html}\\ \hline
			Dutch classroom data & \cite{sewell2015latent} & 2.2.2&\footnotesize\url{http://visone.info/wiki/index.php/Knecht_Classroom_(data)}\\  \hline
			Cosponsorship data & \cite{sewell2015latent} & 2.2.2 &\footnotesize\url{http://jhfowler.ucsd.edu/cosponsorship.htm}\\  \hline
			Friends and Family data & \cite{sewell2016latent} & 2.2.2 & \footnotesize\url{http://realitycommons.media.mit.edu/friendsdataset.html}\\\hline
			World trade data & \cite{sewell2016latent} & 2.2.2 & \footnotesize\url{http://www.economicswebinstitute.org/worldtrade.htm}\\\hline
			Irish companies data & \cite{friel2016interlocking} & 2.2.3 &\footnotesize Irish Stock Exchange and Irish Companies Registration Office websites\\ \hline
			International commerce data & \cite{ward2007persistent} & 2.3.1 & \footnotesize various sources illustrated in Appendix A \cite{ward2007persistent} \\\hline
			Militarized interstate disputes data &\cite{ward2007disputes} & 2.3.1 & \footnotesize various sources illustrated in Appendix A \cite{ward2007disputes} \\\hline
			World trade network & \cite{ward2013gravity} & 2.3.1 & \footnotesize various sources illustrated in Section 3.4 \cite{ward2013gravity}  \\\hline
			National Stock Market Indices & \cite{durante2014} & 2.3.2 & \footnotesize \url{http://finance.yahoo.com/}\\\hline
			Financial Network & \cite{durante2014bayesian} & 2.3.2 & \footnotesize \url{http://finance.yahoo.com/} and \url{http://www.gdeltproject.org/data.html}\\\hline
			Italian soccer Championship data & \cite{durante2014bayesian2} & 2.3.2 & \footnotesize not available\\\hline
			International relations data & \cite{hoff2015multilinear,minhas2016new} & 2.3.3 &\footnotesize\url{https://dataverse.harvard.edu/dataverse/icews}\\\hline
			Southern women data & \cite{yang-etal-2011} & 3.2.1 & \footnotesize Freeman \cite{freeman-2003} \\\hline
			Blog (NEC Labs) data & \cite{yang-etal-2011} & 3.2.1 & \footnotesize Chi et al. \cite{chi-etal-2007}; Lin et al. \cite{lin-etal-2008} \\\hline
			Paper co-authorship (DBLP) data & \cite{yang-etal-2011} & 3.2.1 & \footnotesize Asur et al. \cite{asur-parthasarathy-2007}; Lin et al. \cite{lin-etal-2008} \\\hline
			MIT reality mining data & \cite{xu-hero-2014} & 3.2.2 & \footnotesize Eagle et al. \cite{eagle-etal-2009} \\\hline
			French high school data & \cite{matias-miele-2017} & 3.2.2 & \footnotesize\url{http://www.sociopatterns.org} \\\hline
			Migratory birds (sparrows) data & \cite{matias-miele-2017} & 3.2.2 & \footnotesize Shizuka et al. \cite{shizuka-etal-2014} \url{http://datadryad.org/resource/doi:10.5061/dryad.d3m85} \\\hline
			Indian equids (onagers) data & \cite{matias-miele-2017} & 3.2.2 & \footnotesize Rubenstein et al. \cite{rubenstein-etal-2015}  \url{http://datadryad.org/resource/doi:10.5061/dryad.q660q} \\\hline
			Sampson's monk data & \cite{xing-etal-2010} & 3.2.3 & \footnotesize \url{https://networkdata.ics.uci.edu/netdata/html/sampson.html}  \\\hline
			Enron email data & \cite{xu-hero-2014,xing-etal-2010,ho-etal-2011} & 3.2.2 / 3.2.3 & \footnotesize \url{https://snap.stanford.edu/data/email-Enron.html} \\\hline
			Drosophila melanogaster gene network & \cite{xing-etal-2010} & 3.2.3 & \footnotesize Arbeitman et al. \cite{arbeitman-etal-2002} \\\hline
			US congress voting data & \cite{ho-etal-2011} & 3.2.3 & \footnotesize\url{https://www.senate.gov}\\\hline
		\end{tabular}}
	\end{center}
\end{document}